# Compact Polyhedra of Cubic Symmetry: Geometrical Analysis and Classification

## Klaus E. Hermann

Inorg. Chem. Dept., Fritz-Haber-Institut der Max-Planck-Gesellschaft, Faradayweg 4-6, 14195 Berlin, Germany.

## Abstract

Compact polyhedra of cubic point symmetry $O_h$, exhibit surfaces of planar sections (facets) characterized by normal vector families $\{abc\}$ with up to 48 members each, compatible with $O_h$ symmetry. We focus first on polyhedra confined by facets with normal vectors of one family, $\{100\}$, $\{110\}$, and $\{111\}$, separately. This yields generic polyhedra which serve for the definition of general compact polyhedra as intersections of the three generic species. Their structural properties, such as shape, size, volume, and surface facets, are found to be described by only three polyhedral structure parameters $A$, $B$, $C$. In addition, we examine compact polyhedra exhibiting facets defined by normal vectors of only one general $\{abc\}$ family resulting in up to 48 facets. These polyhedra can be described by four polyhedral structure parameters $Q$, $a$, $b$, $c$. Geometrical properties of all polyhedra are discussed in analytical detail and substantiated by visualization of characteristic examples. Corresponding relationships can also be used to classify shapes and estimate sizes of real compact metal nanoparticles observed e.g. in electron microscopy experiments.

# 1. Introduction

The shape of polyhedral bodies has been studied in great detail since ancient times and probably all mathematical details have been worked out so far [1]. More recently, this subject has attracted new interest in connection with crystalline nanoparticles (NPs) [2-5]. These particles come in many sizes and polyhedral shapes. They have been explored in a large number of experimental and theoretical studies due to their exciting physical and chemical properties which can deviate from those of corresponding bulk material [2-4]. Examples are applications in medicine [6] or in catalytic chemistry where metal nanoparticles have become ubiquitous [7, 8].

While physical and chemical properties of NPs are well known their geometry has, in many cases, not been explored in quantitative detail. Many metal NPs have been shown in experiments to exhibit shapes reflecting cubic $O_h$ symmetry, which can be associated with compact sections of cubic bulk crystal structures [9, 10]. However, further details as to exact polyhedral particle shapes is missing. Here the analysis of ideal polyhedra with cubic $O_h$ symmetry can be helpful to obtain insight into possible NP shapes and their classification.

In the present work, extending previous theoretical analyses [5, 11], we examine geometrical details of ideal polyhedra of cubic $O_h$ symmetry confined by planar facets. The facets are assumed to form finite sections of planes with normal vectors along (1, 0, 0), (1, 1, 0), and (1, 1, 1) directions and their symmetry equivalents in Cartesian coordinates. These directions are suggested by NPs with cubic lattice where possible facets representing high density monolayers of the lattice, {$hkl$} families {100}, {110}, and {111} [10], seem to be energetically preferred. The analysis reveals different types of generic polyhedra which can serve for the definition of general polyhedra described as intersections of corresponding generic species. Their structural properties, such as shape, size, and surface facets, are shown to be determined by only three parameters $A$, $B$, $C$ (polyhedral structure parameters). Further, structural properties of compact polyhedra of $O_h$ symmetry, confined by facets with normal vectors of one general {$abc$} family, are shown to be described by four polyhedral structure parameters $Q$, $a$, $b$, $c$. All results of the present polyhedra are discussed in analytical and numerical detail with visualization [12] of characteristic examples. The sections are structured identically and presented in parts with very similar phrasing to enable easy comparison. These results allow an easy classification of all corresponding polyhedra which may be used as a repository for structures of real compact NPs with internal cubic lattice observed by experiment.

## 2. Formalism and Discussion

We consider compact polyhedra of central **$O_h$ symmetry** confined by finite sections of planes (facets) which can be described by normal vectors $\underline{e}$ and facet distances $R_f$ from the polyhedral center, resulting in facet vectors

$$\underline{R}_f = R_f\, \underline{e} = R_f\, (x, y, z)\,, \qquad x^2 + y^2 + z^2 = 1 \qquad (1)$$

with $\underline{e}$ described in Cartesian coordinates $x, y, z$ relative to the polyhedral center $\underline{o}$ set to $\underline{o} = (0, 0, 0)$. Due to the overall $O_h$ symmetry each facet vector $\underline{R}_f$ implies a number of symmetry equivalents $\underline{R}_f'$ originating from all $O_h$ symmetry operations applied to $\underline{R}_f$, which, together with $\underline{R}_f$, form a family of symmetry equivalent facet vectors denoted $\{\underline{R}_f\}$. As an example, applying all 48 $O_h$ symmetry operations to a facet normal vector $e = (1, 0, 0)$ yields a vector family with 6 members

$$\{\underline{e}\} = \{(\pm 1, 0, 0),\, (0, \pm 1, 0),\, (0, 0, \pm 1)\} \stackrel{\text{def}}{=} \{100\} \qquad (2a)$$

In analogy, the facet normal vector $e = 1/\sqrt{2}\,(1, 1, 0)$ yields a vector family with 12 members

$$\{\underline{e}\} = \{1/\sqrt{2}\,(\pm 1, \pm 1, 0),\, 1/\sqrt{2}\,(\pm 1, 0, \pm 1),\, 1/\sqrt{2}\,(0, \pm 1, \pm 1)\} \stackrel{\text{def}}{=} \{110\} \qquad (2b)$$

and the facet normal vector $e = 1/\sqrt{3}\,(1, 1, 1)$ results to a vector family with 8 members

$$\{\underline{e}\} = \{1/\sqrt{3}\,(\pm 1, \pm 1, \pm 1) \stackrel{\text{def}}{=} \{111\} \qquad (2c)$$

In the following we focus on polyhedra of central $O_h$ symmetry whose facets are described by the three facet vector families $\{100\}, \{110\}$, and $\{111\}$ defined in (2) together with corresponding facet distances $R_{\{100\}}$, $R_{\{110\}}$, and $R_{\{111\}}$, of the three families. Due to the $O_h$ symmetry the polyhedral facets always appear as parallel pairs which suggests **polyhedral diameters**

$$D_{\{100\}} = 2\,R_{\{100\}}\,, \qquad D_{\{110\}} = 2\,R_{\{100\}}\,, \qquad D_{\{111\}} = 2\,R_{\{100\}} \qquad (3)$$

where we define polyhedral parameters for convenience in subsequent relations by

$$D_{\{100\}} = A\,, \qquad D_{\{110\}} = B/\sqrt{2}\,, \qquad D_{\{111\}} = C/\sqrt{3} \qquad (4)$$

Thus the polyhedra can be denoted **P(A, B, C)** in the most general case. If a facet type does not appear at the polyhedron surface the corresponding parameter value $A$, $B$, or $C$ may be ignored and will be replaced by a minus sign. As an example, a polyhedron with only $\{100\}$ and $\{110\}$ facets is denoted P(A, B, -). These notations will be used in the following discussion.

## 2.1. Generic Polyhedra

Generic polyhedra of $O_h$ symmetry are confined by facets with orientations of only one $\{abc\}$ vector family i.e. only $\{100\}$, $\{110\}$, or $\{111\}$ facets. These polyhedra

(a) **P(A, -, -)** polyhedra are confined by all 6 $\{100\}$ planes with distances $D_{\{100\}} = A$ between parallel confining planes. This yields 6 $\{100\}$ facets describing a **cubic** polyhedron, see Fig. 1.

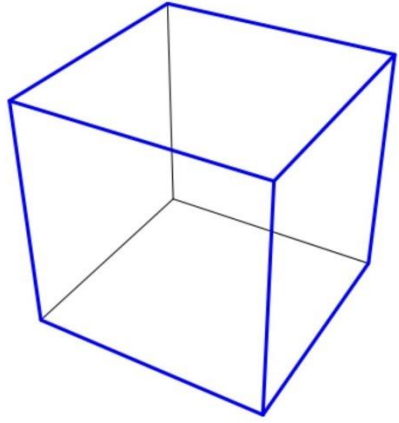

**Figure 1.** Sketch of generic cubic polyhedron with front facets in blue and back facets in black.

The **{100} facets** are square with 4 <100> edges of length $A$.

The 8 polyhedral corners are described by vectors $\underline{c}$ relative to the center where in Cartesian coordinates

$$\underline{r} = 1/2\,(\pm A, \pm A, \pm A) \qquad (5)$$

The largest distance from the polyhedral center to its surface (edges) along <$abc$> directions, $s_{<abc>}$, is given by

$$s_{<100>}(A, -, -) = A/2 \qquad (6a)$$
$$s_{<110>}(A, -, -) = \sqrt{2}\,A/2 \qquad (6b)$$
$$s_{<111>}(A, -, -) = \sqrt{3}\,A/2 \qquad (6c)$$

The polyhedron volume, $V(A, -, -)$, and its surface area, $F(A, -, -)$, (sum over all facet areas), are given by

$$V(A, -, -) = A^3 \qquad (7)$$
$$F(A, -, -) = 6A^2 \qquad (8)$$

**(b)** **P(-, B, -)** polyhedra are confined by all 12 {110} planes with distances $D_{\{110\}} = B/\sqrt{2}$ between parallel confining planes. This yields 12 {110} facets describing a **rhombohedral** polyhedron, see Fig. 2.

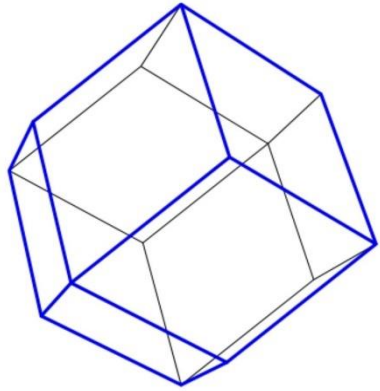

**Figure 2.** Sketch of generic rhombohedral polyhedron with front facets in blue and back facets in black.

The **{110} facets** are rhombic with <111> edges of length $\sqrt{3}\,B/4$. Thus, the polyhedron can be described as a rhombic dodecahedron reminding of the shape of Wigner-Seitz cells of the face-centered cubic crystal lattice [20].

The 14 polyhedral corners are described by vectors $\underline{c}$ relative to the center, falling in two groups of 8 and 6 corners each, where in Cartesian coordinates

$\underline{c} = 1/4\,(\pm B, \pm B, \pm B)$,

$\underline{c} = 1/2\,(\pm B, 0, 0)$,    $= 1/2\,(0, \pm B, 0)$,    $= 1/2\,(0, 0, \pm B)$    (9)

The largest distance from the polyhedral center to its surface along $<abc>$ directions, $s_{<abc>}$, is given by

$s_{<100>}(-, B, -) = B/2$    (10a)

$s_{<110>}(-, B, -) = \sqrt{2}\,B/4$    (10b)

$s_{<111>}(-, B, -) = \sqrt{3}\,B/4$    (10c)

The polyhedron volume, $V(-, B, -)$, and its surface area, $F(-, B, -)$, (sum over all facet areas), are given by

$V(-, B, -) = B^3/4$    (11)

$F(-, B, -) = 3/\sqrt{2}\,B^2$    (12)

(c) **P(-, -, C)** polyhedra are confined by all 8 {111} planes with distances $D_{\{111\}} = C/\sqrt{3}$ between parallel confining planes. This yields 8 {111} facets describing an **octahedral** polyhedron, see Fig. 3.

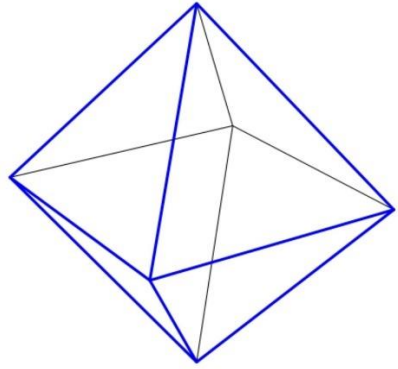

**Figure 3.** Sketch of generic octahedral polyhedron with front facets in blue and back facets in black.

The **{111} facets** are triangular with three <110> edges of length $\sqrt{2}\, C/2$.

The 6 polyhedral corners are described by vectors $\underline{c}$ relative to the center where in Cartesian coordinates

$$\underline{c} = 1/2\,(\pm C, 0, 0)\,, \quad = 1/2\,(0, \pm C, 0)\,, \quad = 1/2\,(0, 0, \pm C) \tag{13}$$

The largest distance from the MP center to its surface along <abc> directions, $s_{<abc>}$, is given by

$$s_{<100>}(-, -, C) = C/2 \tag{14a}$$
$$s_{<110>}(-, -, C) = \sqrt{2}\, C/4 \tag{14b}$$
$$s_{<111>}(-, -, C) = \sqrt{3}\, C/6 \tag{14c}$$

These quantities will be used in Secs. 2.2.

The polyhedron volume, $V(-, -, C)$, and its surface area, $F(-, -, C)$, (sum over all facet areas), are given by

$$V(-, -, C) = C^3/6 \tag{15}$$
$$F(-, -, C) = \sqrt{3}\, C^2 \tag{16}$$

Table 1 of Supplement S.2 collects types and shapes of all generic P(A, -, -), (-, B, -), (-, -, C) polyhedra.

## 2.2. Non-generic Polyhedra

Non-generic polyhedra of $O_h$ symmetry show facets with orientations of more than one $\{abc\}$ vector family. This can be considered as combining confinements of the corresponding generic polyhedra, discussed in Sec. 2.1, sharing their symmetry center. Thus, non-generic polyhedra are mutual intersections of more than one generic polyhedron, where one cuts corners and edges from the other(s) to form additional facets. Here we discuss non-generic polyhedra **P(*A*, *B*, *C*)**, which combine constraints of up to three generic polyhedra, cubic P(*A*, -, -), rhombohedral P(-, *B*, -), and octahedral P(-, -, *C*). Thus, they allow {100}, {110}, as well as {111} facets. Clearly, the corresponding polyhedral parameters *A*, *B*, *C* depend on each other and determine the overall polyhedron shape. In particular, if a participating generic polyhedron encloses another participant it will not contribute to the overall polyhedron shape. Thus, the respective $\{abc\}$ facets will not appear at the surface of the non-generic polyhedron. In the following, we consider the three types of non-generic polyhedra, which combine constraints due to two generic polyhedra (Secs. 2.2.1-3), before we discuss the most general case of P(*A*, *B*, *C*) MPs in Sec. 2.2.4.

## 2.2.1. Non-generic Polyhedra P(*A*, *B*, -)

Non-generic **cubo-rhombic** polyhedra, denoted **P(*A*, *B*, -)**, are confined by facets of the two generic polyhedra, P(*A*, -, -) (cubic) and P(-, *B*, -) (rhombohedral), see Fig. 4. If the edges of the cubic polyhedron P(*A*, -, -) lie inside the rhombohedral polyhedron P(-, *B*, -), the resulting combination, P(*A*, *B*, -), will be generic cubic. This requires

$$s_{<110>}(A, -, -) \leq s_{<110>}(-, B, -) \tag{17}$$

and, with (6), (10), leads to

$$B \geq 2A \tag{18}$$

On the other hand, if the corners of the rhombohedral polyhedron P(-, *B*, -) lie inside the cubic polyhedron P(*A*, -, -), the resulting combination P(*A*, *B*, -) will be generic rhombohedral. This requires

$$s_{<100>}(-, B, -) \leq s_{<100>}(A, -, -) \tag{19}$$

and, with (6), (10), leads to

$$B \leq A \tag{20}$$

Thus, the two generic polyhedra intersect and yield a polyhedron P(*A*, *B*, -) with both {100} and {110} facets only for polyhedral parameters *A*, *B* with

$$A < B < 2A \qquad 1 < B/A < 2 \qquad (21)$$

while P(A, B, -) is generic cubic for $B \geq 2A$ and generic rhombohedral for $B \leq A$. Further, generic cubic and rhombohedral polyhedra can be described by P(A, B, -) where

$$P(A, -, -) = P(A, B = 2A, -) \qquad \text{(cubic)} \qquad (22a)$$
$$P(-, B, -) = P(A = B, B, -) \qquad \text{(rhombohedral)} \qquad (22b)$$

The surfaces of cubo-rhombic polyhedra P(A, B, -) exhibit 6 {100} facets and 12 {110} facets, see Fig. 4.

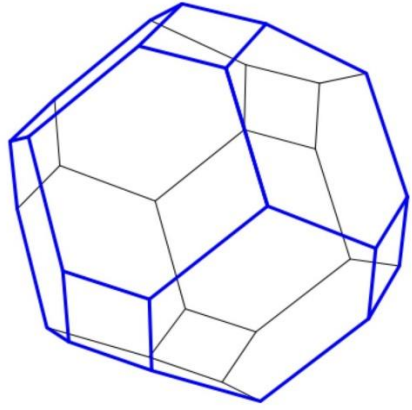

**Figure 4.** Sketch of cubo-rhombic polyhedron ($B/A = 1.22$) with front facets in blue and back facets in black.

The **{100} facets** are square with 4 <100> edges of length $(B - A)$.

The **{110} facets** are hexagonal with 4 <111> edges of length $\sqrt{3}/4\,(2A - B)$ and two <100> edges of length $(B - A)$.

The 32 polyhedral corners are described by vectors $\underline{c}$ relative to the center, falling in two groups of 8 and 24 corners each, where in Cartesian coordinates

$$\underline{c} = 1/2\,(\pm A, \pm H, \pm H)\,, \quad = 1/2\,(\pm H, \pm A, \pm H)\,, \quad = 1/2\,(\pm H, \pm H, \pm A) \qquad (23)$$
$$\underline{c} = 1/4\,(\pm B, \pm B, \pm B)\,, \quad H = B - A$$

The largest distance from the polyhedral center to its surface along <abc> directions, $s_{<abc>}$, is given by

$$s_{<100>}(A, B, -) = A/2 \qquad (24a)$$
$$s_{<110>}(A, B, -) = \sqrt{2}\,B/4 \qquad (24b)$$
$$s_{<111>}(A, B, -) = \sqrt{3}\,B/4 \qquad (24c)$$

The polyhedron volume, **$V(A, B, -)$**, and its surface area, **$F(A, B, -)$**, (sum over all facet areas), are given by

$$V(A, B, -) = B^3/4 - (B - A)^3 \qquad (25)$$

$$F(A, B, -) = 6(B - A)^2 + 3/\sqrt{2}\,(2A - B)(3B - 2A) \qquad (26)$$

A classification of P($A$, $B$, -) polyhedra for any $A$, $B$ combination is given by Table 2 of Supplement S.2.

### 2.2.2. Non-generic Polyhedra P($A$, -, $C$)

Non-generic **cubo-octahedral** polyhedra, denoted **P($A$, -, $C$)**, are confined by facets of the two generic polyhedra, P($A$, -, -) (cubic) and P(-, -, $C$) (octahedral), see Fig. A5. If the corners of the cubic polyhedron P($A$, -, -) lie inside the octahedral polyhedron P(-, -, $C$), the resulting combination P($A$, -, $C$) will be generic cubic. This requires

$$s_{<111>}(A, -, -) \leq s_{<111>}(-, -, C) \qquad (27)$$

and with (6), (14), leads to

$$C \geq 3A \qquad (28)$$

On the other hand, if the corners of the octahedral polyhedron P(-, -, $C$) lie inside the cubic polyhedron P($A$, -, -), the resulting combination P($A$, -, $C$) will be generic octahedral. This requires

$$s_{<100>}(-, -, C) \leq s_{<100>}(A, -, -) \qquad (29)$$

and with (D.2), (D.4), leads to

$$C \leq A \qquad (30)$$

Thus, the two generic polyhedra intersect and yield a polyhedron P($A$, -, $C$) with both {100} and {111} facets only for polyhedral parameters $A$, $C$ with

$$A < C < 3A \qquad 1 < C/A < 3 \qquad (31)$$

while P($A$, -, $C$) is generic cubic for $C \geq 3A$ and generic octahedral for $C \leq A$. Further, generic cubic and octahedral cub polyhedra can be described by P($A$, -, $C$) where

$$P(A, -, -) = P(A, -, C = 3A) \qquad \text{(cubic)} \qquad (32a)$$

$$P(-, -, C) = P(A = C, -, C) \qquad \text{(octahedral)} \qquad (32b)$$

The surfaces of cubo-octahedral polyhedra P($A$, -, $C$) include 6 {100} and 8 {111} facets, see Figs. 5, 6. Amongst the intersecting species according to (31) we can distinguish between **truncated octahedral** polyhedra where $A < C < 2A$ and **truncated cubic** polyhedra for $2A < C < 3A$, with **cuboctahedral** polyhedra for $C = 2A$ separating. This will be discussed in the following.

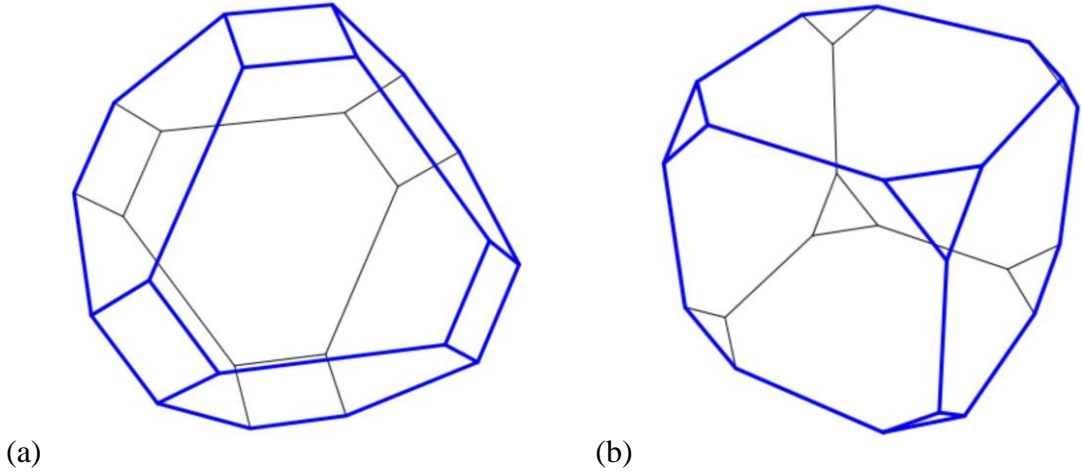

(a)        (b)

**Figure 5.** Sketch of cubo-octahedral polyhedra with front facets in blue and back facets in black, (a) truncated octahedral ($C/A = 1.33$), (b) truncated cubic type ($C/A = 2.6$).

**Truncated octahedral** polyhedra ($A < C < 2A$), Fig. 5a, can be characterized by their facets as follows:

The **{100} facets** are square with 4 <110> edges of length $\sqrt{2}\,(C - A)/2$.

The **{111} facets** are hexagonal with <110> edges of alternating lengths $\sqrt{2}\,(C - A)/2$ and $\sqrt{2}\,(2A - C)/2$. For $C = 3/2\,A$ all 6 edge lengths are equal leading to regular hexagonal {111} facets. As a result, the polyhedron is reminiscent of the shape of Wigner-Seitz cells of the body-centered cubic crystal lattice [20].

The 24 polyhedral corners are described by vectors $\underline{c}$ relative to the center where in Cartesian coordinates

$$\underline{c} = 1/2\,(\pm A, \pm H, 0)\,, \quad = 1/2\,(\pm H, \pm A, 0)\,, \quad = 1/2\,(\pm H, 0, \pm A)\,,$$
$$\underline{c} = 1/2\,(\pm A, 0, \pm H)\,, \quad = 1/2\,(0, \pm A, \pm H)\,, \quad = 1/2\,(0, \pm H, \pm A) \tag{33}$$
$$H = C - A$$

The largest distance from the polyhedral center to its surface along <abc> directions, $s_{<abc>}$, is given by

$$s_{<100>}(A, -, C) = A/2 \tag{34a}$$
$$s_{<110>}(A, -, C) = \sqrt{2}\,C/4 \tag{34b}$$
$$s_{<111>}(A, -, C) = \sqrt{3}\,C/6 \tag{34c}$$

The polyhedron volume, $V(A, -, C)$, and its surface area, $F(A, -, C)$, (sum over all facet areas), are given by

$$V(A, -, C) = C^3/6 - 1/2(C - A)^3 \tag{35}$$

$$F(A, -, C) = 3(C - A)^2 + \sqrt{3}\,[C^2 - 3(C - A)^2] \tag{36}$$

**Truncated cubic** polyhedron ($2A < C < 3A$), Fig. 5b, can be characterized by their facets as follows:

The **{100} facets** are octagonal with alternating edges, 4 <100> of length $(C - 2A)$ and 4 <110> of length $\sqrt{2}\,(3A - C)/2$.

The **{111} facets** are triangular with <110> edges of length $\sqrt{2}\,(3A - C)/2$.

The 24 polyhedral corners are described by vectors $\underline{c}$ relative to the center where in Cartesian coordinates

$$\underline{c} = 1/2\,(\pm H, \pm A, \pm A), \quad = 1/2\,(\pm A, \pm H, \pm A), \quad = 1/2\,(\pm A, \pm A, \pm H) \tag{37}$$

$$H = C - 2A$$

The largest distance from the polyhedral center to its surface along <abc> directions, $s_{<abc>}$, is given by

$$s_{<100>}(A, -, C) = A/2 \tag{38a}$$

$$s_{<110>}(A, -, C) = \sqrt{2}\,A/2 \tag{38b}$$

$$s_{<111>}(A, -, C) = \sqrt{3}\,C/6 \tag{38c}$$

The polyhedron volume, $V(A, -, C)$, and its surface area, $F(A, -, C)$, (sum over all facet areas), are given by

$$V(A, -, C) = A^3 - 1/6\,(3A - C)^3 \tag{39}$$

$$F(A, -, C) = 6A^2 - (3 - \sqrt{3})(3A - C)^2 \tag{40}$$

There are polyhedra which can be assigned to both truncated cubic and truncated octahedral type, the **cuboctahedral** polyhedra $P(A, -, C)$, defined by $C = 2A$ These polyhedra exhibit 6 {100} and 8 {111} facets, see Fig. 6. All **{100} facets** are square with 4 <110> edges of length $A/\sqrt{2}$ while all **{111} facets** are triangular with three <110> edges of length $A/\sqrt{2}$ shared with those of the {100} facets.

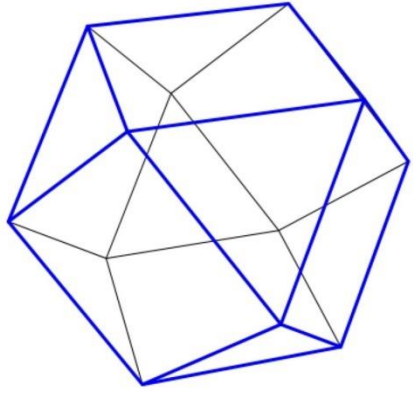

**Figure 6.** Sketch of cuboctahedral polyhedron ($C/A = 2$) with front facets in blue and back facets in black.

The 12 polyhedral corners are described by vectors $\underline{c}$ relative to the center where in Cartesian coordinates

$$\underline{c} = 1/2\,(0, \pm A, \pm A),\quad = 1/2\,(\pm A, 0, \pm A),\quad = 1/2\,(\pm A, \pm A, 0) \tag{41}$$

The largest distance from the polyhedral center to its surface along $<abc>$ directions, $s_{<abc>}$, is given by

$$s_{<100>}(A, -, C) = A/2 \tag{42a}$$
$$s_{<110>}(A, -, C) = \sqrt{2}\,A/2 \tag{42b}$$
$$s_{<111>}(A, -, C) = \sqrt{3}\,A/3 \tag{42c}$$

The polyhedron volume, $V(A, B, -)$, and its surface area, $F(A, B, -)$, (sum over all facet areas), are given by

$$V(A, -, C) = 5/6\,A^3 \tag{43}$$
$$F(A, -, C) = (3 + \sqrt{3})\,A^2 \tag{44}$$

A classification of P($A$, -, $C$) polyhedra types for any $A$, $C$ combination is given by Table 3 of Supplement S.2.

### 2.2.3. Non-generic Polyhedra P(-, *B*, *C*)

Non-generic **rhombo-octahedral** polyhedrons, denoted **P(-, *B*, *C*)**, are confined by facets of the two generic polyhedra, P(-, *B*, -) (rhombohedral) and P(-, -, *C*) (octahedral), see Fig. 7. If the corners of the rhombohedral polyhedron P(-, *B*, -) lie inside the octahedral polyhedron P(-, -, *C*), the resulting combination P(-, *B*, *C*) will be generic rhombohedral. This requires

$$s_{<111>}(-, B, -) \leq s_{<111>}(-, -, C) \tag{45}$$

and with (10), (14), leads to

$$C \geq 3/2\, B \tag{46}$$

On the other hand, if the corners of the octahedral polyhedron P(-, -, *C*) lie inside the rhombohedral polyhedron P(-, *B*, -), the resulting combination P(-, *B*, *C*) will be generic octahedral. This requires

$$s_{<100>}(-, -, C) \leq s_{<100>}(-, B, -) \tag{47}$$

and, with (D.3), (D.4), leads to

$$C \leq B \tag{48}$$

Thus, the two generic polyhedra intersect and yield an polyhedron P(-, *B*, *C*) with both {110} and {111} facets only for polyhedral parameters *B*, *C* with

$$B < C < 3/2\, B \qquad 1 < C/B < 3/2 \tag{49}$$

while P(-, *B*, *C*) is generic rhombohedral for $C \geq 3/2$ and generic octahedral for $C \leq B$. Further, generic rhombohedral and octahedral polyhedra can be described by P(-, *B*, *C*) where

$$P(-, B, -) = P(-, B, C = 3/2\, B) \qquad \text{(rhombohedral)} \tag{50a}$$

$$P(-, -, C) = P(-, B = C, C) \qquad \text{(octahedral)} \tag{50b}$$

The surfaces of rhombo-octahedral polyhedrons P(-, *B*, *C*) exhibit 12 {110} and 8 {111} facets, see Fig. 7.

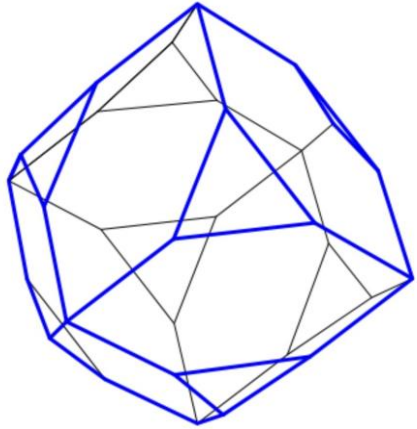

**Figure 7.** Sketch of rhombo-octahedral polyhedron ($C/B = 1.3$) with front facets in blue and back facets in black.

The **{110} facets** are hexagonal with 4 <111> edges of length $(C - B)/2 \sqrt{3}$ and two <110> edges of $(3/2\, B - C) \sqrt{2}$.

The **{111} facets** are triangular with three <110> edges of length $(3/2\, B - C) \sqrt{2}$.

The 30 polyhedral corners are described by vectors $\underline{c}$ relative to the center, falling in two groups of 6 and 24 corners each, where in Cartesian coordinates

$$\underline{c} = 1/2\,(\pm B, 0, 0), \qquad = 1/2\,(0, \pm B, 0), \qquad = 1/2\,(0, 0, \pm B)$$
$$\underline{c} = 1/2\,(\pm H, \pm G, \pm G), \qquad = 1/2\,(\pm G, \pm H, \pm G), \qquad = 1/2\,(\pm G, \pm G, \pm H) \qquad (51)$$
$$H = 2B - C, \qquad G = C - B$$

The largest distance from the polyhedral center to its surface along <abc> directions, $s_{<abc>}$, is given by

$$s_{<100>}(-, B, C) = B/2 \qquad (52a)$$
$$s_{<110>}(-, B, C) = \sqrt{2}\, B/4 \qquad (52b)$$
$$s_{<111>}(-, B, C) = \sqrt{3}\, C/6 \qquad (52c)$$

The polyhedron volume, $V(-, B, C)$, and its surface area, $F(-, B, C)$, (sum over all facet areas), are given by

$$V(-, B, C) = B^3/4 - 1/12\,(3B - 2C)^3 \qquad (53)$$
$$F(-, B, C) = \sqrt{3}\,(3B - 2C)^2 + 3/\sqrt{2}\,[B^2 - (3B - 2C)^2] \qquad (54)$$

A classification of $P(-, B, C)$ polyhedron types for any $B, C$ combination is given by Table 4 of Supplement S.2.

## 2.2.4. General non-generic Polyhedra P(*A*, *B*, *C*)

Non-generic **cubo-rhombo-octahedral** polyhedra, denoted **P(*A*, *B*, *C*)**, are confined by facets of all three generic polyhedra, P(*A*, -, -) (cubic), P(-, *B*, -) (rhombohedral), and P(-, -, *C*) (octahedral). Thus, they can show {100}, {110}, and {111} facets. A general discussion of these polyhedra requires results for generic and non-generic polyhedra, see Secs. 2.1, and 2.2.1-3, as will be detailed in the following.

First, we consider the general notation for generic polyhedra discussed in Sec. 2.1. Cubic polyhedra P(*A*, -, -) are surrounded by smallest rhombohedral polyhedra P(-, *B*, -) if $B = 2A$ and by smallest octahedral polyhedra P(-, -, *K*) if $C = 3A$. This allows a notation

$$P(A, -, -) = P(A, B = 2A, C = 3A) \tag{55}$$

Rhombohedral polyhedra P(-, *B*, -) are surrounded by smallest cubic polyhedra P(*A*, -, -) if $A = B$ and by smallest octahedral polyhedra P(-, -, *C*) if $C = 3/2\, B$. This yields

$$P(-, B, -) = P(A = B, B, C = 3/2\, B) \tag{56}$$

Octahedral polyhedra P(-, -, *C*) are surrounded by smallest cubic polyhedra P(*A*, -, -) if $A = C$ and by smallest rhombohedral polyhedra P(-, *B*, -) if $B = C$. This yields

$$P(-, -, C) = P(A = C, B = C, C) \tag{57}$$

General notations for non-generic cub polyhedra with two facet types, discussed in Secs. 2.2.1-3, are obtained by analogous arguments. Cubo-rhombic polyhedra P(*A*, *B*, -) are surrounded by smallest octahedral polyhedra P(-, -, *C*) if $C = C_a$ with

$$C_a(A, B) = \min(3A, 3/2\, B) = 3/2\, B \tag{58}$$

which allows a notation

$$P(A, B, -) = P(A, B, C = C_a) \tag{59}$$

Cubo-octahedral polyhedra P(*A*, -, *C*) surrounded by smallest rhombohedral polyhedra P(-, *B*, -) if $B = B_a$ with

$$B_a(A, C) = \min(2A, C) \tag{60a}$$
$$= 2A \quad \text{(truncated cubic, } C \geq 2A\text{)} \tag{60b}$$
$$= C \quad \text{(truncated octahedral, } C \leq 2A\text{)} \tag{60c}$$

yielding

$$P(A, -, C) = P(A, B = B_a, C) \tag{61}$$

Rhombo-octahedral polyhedra P(-, $B$, $C$) are surrounded by smallest cubic polyhedra P($A$, -, -) if $A = A_a$ with

$$A_a(B, C) = \min(B, C) = B \tag{62}$$

yielding

$$P(-, B, C) = P(A = A_a, B, C) \tag{63}$$

In the most general case of a P($A$, $B$, $C$) polyhedron, we start from a cubo-rhombic polyhedron, P($A$, $B$, -). Then we add constraints of a generic octahedral polyhedron, P(-, -, $C$), to yield the cubo-rhombo-octahedral polyhedron P($A$, $B$, $C$). Here we can distinguish three different scanarios depending on the geometry of the initial polyhedron described by relationships between structure parameters $A$ and $B$, i.e. $B \leq A$, $A < B < 2A$, or $B \geq 2A$, as discussed in the following.

A classification of P($A$, $B$, $C$) polyhedron types for any $A$, $B$, $C$ combination is given by Table 5 of Supplement S.2.

## 2.2.4.1. P($A$, $B$, $C$) by truncating cubo-rhombic P($A$, $B$, -)

The initial polyhedron **P($A$, $B$, -)** with $A < B < 2A$ is a true cubo-rhombic polyhedron. As a result, imposing constraints of a generic octahedral polyhedron, P(-, -, $C$), can yield a true cubo-rhombo-octahedral polyhedron P($A$, $B$, $C$). This requires, according to the discussion above, $C$ values below $C_a$. Here we distinguish 5 different ranges of parameter $C$, defined by separating values $C_a \geq C_b \geq C_c \geq C_d$, where with $C_a$ from (58) (repeated for clarity) and (60)

$$C_a(A, B) = 3/2\, B \qquad C_a(A, B)/B = 3/2 \tag{58}$$
$$C_b(A, B) = 2B - A \qquad C_b(A, B)/B = 2 - A/B \tag{64}$$
$$C_c(A, B) = B \qquad C_c(A, B)/B = 1 \tag{65}$$
$$C_d(A, B) = A \qquad C_d(A, B)/B = A/B \tag{66}$$

The ranges are illustrated in Fig. 8 for the cubo-rhombic polyhedron P($A$, $B$, $C_a$).

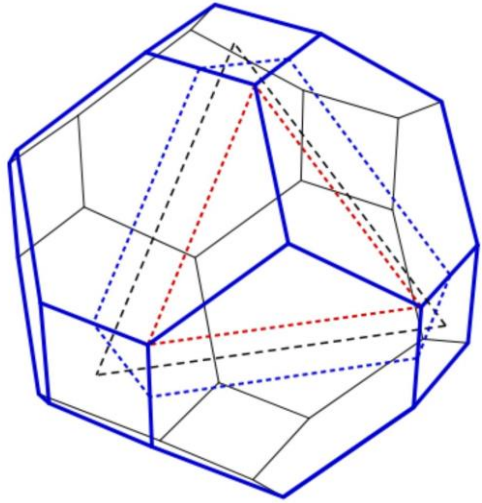

**Figure 8.** Sketch of cubo-rhombic polyhedron P($A$, $B$, $C_a$) ($B/A = 1.25$, $C/B = 1.5$) with front facets in blue and back facets in black. Cuts along {111} facets corresponding to cubo-rhombo-octahedral P($A$, $B$, $C_b$) ($B/A = 1.25$, $C/B = 1.2$, red dashed triangle), cubo-octahedral P($A$, $B$, $C_c$) ($B/A = 1.25$, $C/B = 1.0$, blue dashed hexagon), and octahedral P($A$, $B$, $C_d$) ($B/A = 1$, $C/B = 1.0$, black dashed triangle), are added for illustration.

**Outer $C$ range** of P($A$, $B$, $C$) where with (58)

$$C \geq C_a \qquad\qquad C/B \geq 3/2 \qquad\qquad (67)$$

For these $C$ values the polyhedron becomes cubo-rhombic and does not exhibit any {111} facets, see Fig. 8. It is structurally identical with P($A$, $B$, $C_a$) = P($A$, $B$, -) as discussed above and in Sec. 2.2.1 which details all facet shapes, edges, corner coordinates, etc.

**Upper central $C$ range** of P($A$, $B$, $C$) where with (58), (64)

$$C_b \leq C \leq C_a \qquad\qquad 2 - A/B \leq C/B \leq 3/2 \qquad\qquad (68)$$

For these $C$ values the initial P($A$, $B$, $C_a$) polyhedron is capped at its <111> corners forming {111} facets. Altogether, these polyhedra exhibit 6 {100}, 12 {110}, and 8 {111} facets, see Fig. 9.

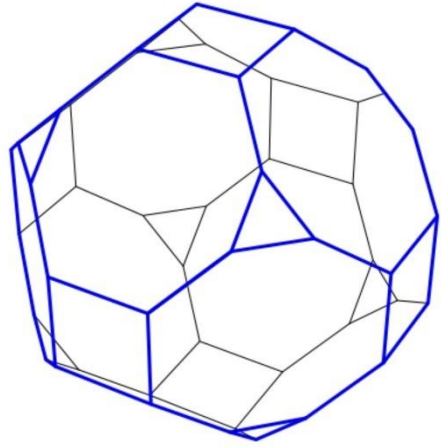

**Figure 9.** Sketch of cubo-rhombo-octahedral polyhedron P(*A*, *B*, *C*) (*B*/*A* = 1.25, *C*/*B* = 1.4) for $C_b < C < C_a$ with front facets in blue and back facets in black.

The **{100} facets** are square with 4 <100> edges of length (*B* - *A*).

The **{110} facets** are octagonal or rectangular (*C* = $C_b$) with two <110> edges of length (3*B* - 2*C*)/√2, two <100> edges of (*B* - *A*), and 4 <111> edges of (*C* - $C_b$)/2 √3.

The **{111} facets** are triangular with three <110> edges of length (3*B* - 2*C*)/√2.

The 48 polyhedral corners are described by vectors $\underline{c}$ relative to the center, falling in two groups of 24 corners each, where in Cartesian coordinates

$$\underline{c} = 1/2\ (\pm A, \pm D, \pm D)\,, \quad = 1/2\ (\pm D, \pm A, \pm D)\,, \quad = 1/2\ (\pm D, \pm D, \pm A)$$
$$\underline{c} = 1/2\ (\pm G, \pm H, \pm H)\,, \quad = 1/2\ (\pm H, \pm G, \pm H)\,, \quad = 1/2\ (\pm H, \pm H, \pm G) \qquad (69)$$
$$D = B - A, \qquad G = 2B - C \qquad H = C - B$$

The largest distance from the polyhedral center to its surface along <*abc*> directions, $s_{<abc>}$, is given by

$$s_{<100>}(A, B, C) = A/2 \qquad (70a)$$
$$s_{<110>}(A, B, C) = \sqrt{2}\ B/4 \qquad (70b)$$
$$s_{<111>}(A, B, C) = \sqrt{3}\ C/6 \qquad (70c)$$

The polyhedron volume, **V**(***A*, *B*, *C***), and its surface area, **F**(***A*, *B*, *C***), (sum over all facet areas), are given by

$$V(A, B, C) = B^3/4 - (B - A)^3 - 1/12\ (3B - 2C)^3 \qquad (71)$$
$$F(A, B, C) = 6(B - A)^2 + \sqrt{3}\ (3B - 2C)^2 + 6\sqrt{2}\ [(2A - B)(C - B) + (A + C - 2B)^2] \qquad (72)$$

**Between** the upper and lower central $C$ range, i.e. for $C = C_b$ ($C/B = 2 - A/B$), the P($A$, $B$, $C$) polyhedron assumes a particular shape, see Fig. 10. Its 6 **{100} facets** are square with 4 <100> edges of length ($B - A$). Its 12 **{110} facets** are rectangular with two <110> edges of length $(2A - B)/\sqrt{2}$ and two <100> edges of ($B - A$). Finally, its 8 **{111} facets** are triangular with three <110> edges of length $(2A - B)/\sqrt{2}$. This is illustrated in Fig. 10.

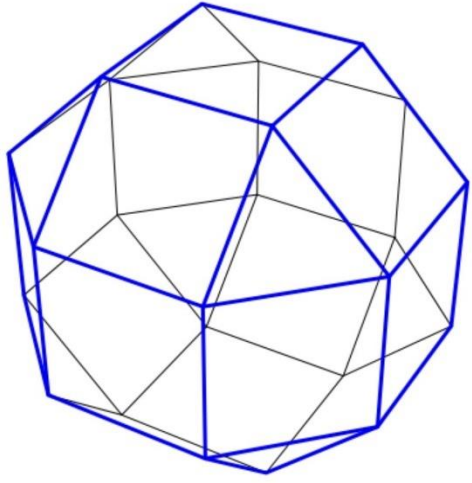

**Figure 10.** Sketch of cubo-rhombo-octahedral polyhedron P($A$, $B$, $C_b$) ($B/A = 1.43$, $C/B = 1.3$) with front facets in blue and back facets in black.

The 24 polyhedral corners are described by vectors $\underline{c}$ relative to the center where in Cartesian coordinates

$$\underline{c} = 1/2\,(\pm A, \pm D, \pm D), \quad = 1/2\,(\pm D, \pm A, \pm D), \quad = 1/2\,(\pm D, \pm D, \pm A) \tag{73}$$
$$D = B - A$$

The largest distance from the polyhedral center to its surface along <$abc$> directions, $s_{<abc>}$, is given by

$$s_{<100>}(A, B, C_b) = A/2 \tag{74a}$$
$$s_{<110>}(A, B, C_b) = \sqrt{2}\,B/4 \tag{74b}$$
$$s_{<111>}(A, B, C_b) = \sqrt{3}/6\,(2B - A) \tag{74c}$$

The polyhedron volume, $V(A, B, C_b)$, and its surface area, $F(A, B, C_b)$, (sum over all facet areas), are given by

$$V(A, B, C_b) = B^3/4 - (B - A)^3 - 1/12\,(2A - B)^3 \tag{75}$$
$$F(A, B, C_b) = 6(B - A)^2 + \sqrt{3}\,(2A - B)^2 + 6\sqrt{2}\,(2A - B)(B - A) \tag{76}$$

**Lower central *C* range** of P(*A*, *B*, *C*) where with (64), (65)

$$C_c \leq C \leq C_b \qquad\qquad 1 \leq C/B \leq 2 - A/B \qquad\qquad (77)$$

For these *C* values the capping of the initial P(*A*, *B*, $C_b$) along the <111> directions is continued to yield hexagonal {111} facets. As before, these polyhedra exhibit 6 {100} facets, 12 {110} facets, and 8 {111} facets, see Fig. 11.

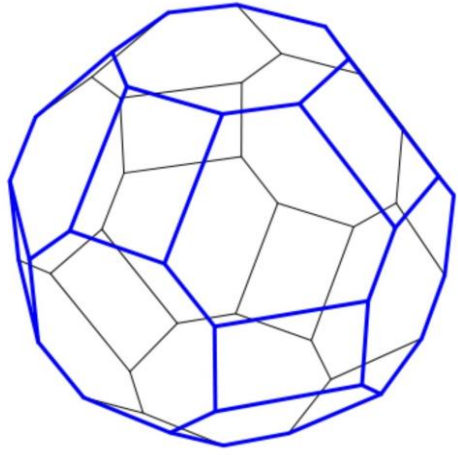

**Figure 11.** Sketch of cubo-rhombo-octahedral polyhedron P(*A*, *B*, *C*) (*B*/*A* = 1.5, *C*/*B* = 1.17) for $C_c < C < C_b$ with front facets in blue and back facets in black.

The **{100} facets** are octagonal with alternating edges, 4 <100> of length (*C* - *B*) and 4 <110> of length (2*B* - *A* - *C*)/√2.

The **{110} facets** are rectangular with two <110> edges of length (2*A* - *B*)/√2 and two <100> edges of length (*C* - *B*).

The **{111} facets** are hexagonal with <110> edges of alternating lengths (2*B* - *A* - *C*)/√2 and (2*A* - *B*)/√2.

The 48 polyhedral corners are described by vectors *c* relative to the center, falling in two groups of 24 corners each, where in Cartesian coordinates

$$\underline{c} = 1/2\ (\pm A, \pm G, \pm H)\ , \qquad = 1/2\ (\pm G, \pm A, \pm H)\ , \qquad = 1/2\ (\pm G, \pm H, \pm A)$$
$$\underline{c} = 1/2\ (\pm A, \pm H, \pm G)\ , \qquad = 1/2\ (\pm H, \pm A, \pm G)\ , \qquad = 1/2\ (\pm H, \pm G, \pm A) \qquad (78)$$
$$G = B - A, \qquad H = C - B$$

The largest distance from the polyhedral center to its surface along <*abc*> directions, $s_{<abc>}$, is given by

$$s_{<100>}(A, B, C) = A/2 \tag{79a}$$
$$s_{<110>}(A, B, C) = \sqrt{2}\,B/4 \tag{79b}$$
$$s_{<111>}(A, B, C) = \sqrt{3}\,C/6 \tag{79c}$$

The polyhedron volume, $V(A, B, C)$, and its surface area, $F(A, B, C)$, (sum over all facet areas), are given by

$$V(A, B, C) = C^3/6 - 1/2\,(C - A)^3 - 1/2\,(C - B)^2\,(6A - 2B - C) \tag{80}$$

$$F(A, B, C) = 3(C - A)^2 - 6(C - B)^2 + 6\sqrt{2}\,(C - B)\,(2A - B)$$
$$+ \sqrt{3}\,[\,(3A - C)^2 - 3(2A - B)^2\,] \tag{81}$$

**Inner $C$ range** of P($A$, $B$, $C$) where with (65), (66), (60)

$$C_d \leq C \leq C_c \qquad\qquad A/B \leq C/B \leq 1 \tag{82}$$

For these $C$ values, the P($A$, $B$, $C$) polyhedron becomes cubo-octahedral (truncated octahedral type) or cuboctahedral and does not exhibit any {110} facets, see Fig. 12. It is structurally identical with P($A$, $B = C$, $C$) = P($A$, -, $C$) = as discussed above and in Sec. 2.2.2 which details all facet shapes, edges, corner coordinates, etc.

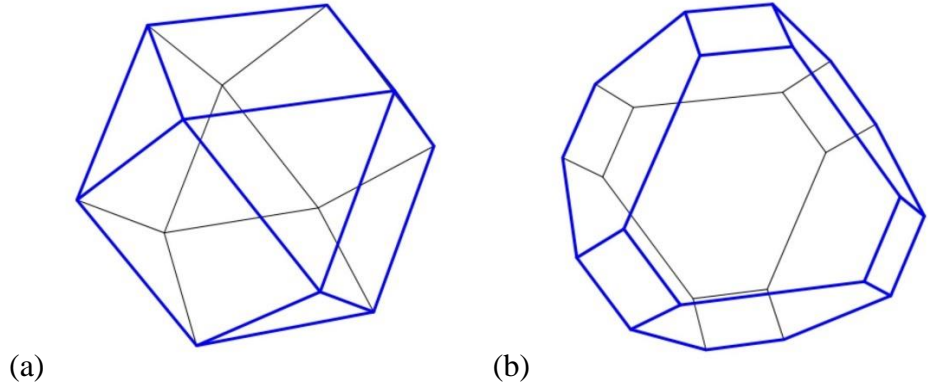

(a)  (b)

**Figure 12.** Sketch of cubo-rhombo-octahedral polyhedra P($A$, $B = C$, $C$) with front facets in blue and back facets in black; (a) cuboctahedral ($B/A = 2$, $C/B = 1$, $C = A$), (b) truncated octahedral ($B/A = 1.333$, $C/B = 1$, $C < 2A$).

**Core $C$ range** of P($A$, $B$, $C$) where with (66)

$$C \leq C_d \qquad\qquad C/B \leq A/B \tag{83}$$

For these $C$ values, the P($A$, $B$, $C$) polyhedron becomes octahedral and exhibits only {111} facets, see Fig. 3. It is structurally identical with P($A = C$, $B = C$, $C$) = P(-, -, $C$) as discussed in Sec. 2.2.4 above and in Sec. 2.1 which details all facet shapes, edges, corner coordinates, etc.

## 2.2.4.2. P(*A*, *B*, *C*) by truncating cubic P(*A*, -, -)

The initial polyhedron **P(*A*, *B*, -) with *B* ≥ 2*A*** is isomorphous to cubic P(*A*, *B* = 2*A*, -) = P(*A*, -, -) discussed in Sec. 2. 1. As a result, imposing constraints of a generic octahedral polyhedron, P(-, -, *C*), can yield a cubo-octahedral polyhedron P(*A*, -, *C*). This requires, according to the discussion above with (58), *C* values below $C_a$ = 3*A*. Here we distinguish 4 different ranges of parameter *C*, defined by separating values $C_a \geq C_c \geq C_d$, where with (58), (65), and (66)

$$C_a(A, B) = 3A \tag{84}$$

$$C_c(A, B) = 2A \tag{85}$$

$$C_d(A, B) = A \tag{86}$$

The ranges are illustrated in Fig. 13 for the cubic polyhedron P(*A*, -, $C_a$).

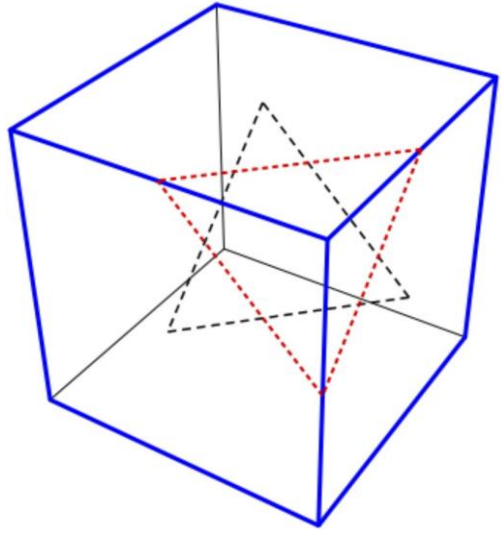

**Figure 13.** Sketch of cubic polyhedron P(*A*, -, $C_a$) (*B*/*A* = 2, *C*/*B* = 1.5) with front facets in blue and back facets in black. Cuts along {111} facets corresponding to cuboctahedral P(*A*, -, $C_c$) (*B*/*A* = 2, *C*/*B* = 1, red dashed triangle) and octahedral P(*A*, -, $C_d$) (*B*/*A* = 1, *C*/*B* = 1, black dashed triangle), are added for illustration.

**Outer *C* range** of P(*A*, -, *C*) where with (84)

$$C \geq C_a \qquad C/A \geq 3 \tag{87}$$

For these *C* values the polyhedron becomes cubic and does not exhibit any {111} facets, see Fig. 13. It is structurally identical with P(*A*, -, $C_a$) = P(*A*, -, -) as discussed above and in Sec. 2.2.2 which details all facet shapes, edges, corner coordinates, etc.

**Upper central $C$ range** of P($A$, -, $C$) where with (84), (85)

$$C_c \leq C \leq C_a \qquad\qquad 2 \leq C/A \leq 3 \qquad\qquad (88)$$

For these $C$ values the initial cubic P($A$, -, $C_a$) polyhedron is capped at its <111> corners forming {111} in addition to {100} facets and yielding a cubo-octahedral polyhedron of truncated cubic type as discussed in Sec. 2.2.2 which details all facet shapes, edges, corner coordinates, etc.

**Between** the upper and lower central $C$ range, i.e. for $C = C_c$ ($C/A = 2$), the P($A$, -, $C$) polyhedron assumes a cuboctahedral shape, see Fig. 6. Its 6 {100} facets are square while all {111} facets are triangular as discussed in Sec. 2.2.2 which details all facet shapes, edges, corner coordinates, etc.

**Lower central $C$ range** of P($A$, -, $C$) where with (85), (86)

$$C_d \leq C \leq C_c \qquad\qquad 1 \leq C/A \leq 2 \qquad\qquad (89)$$

For these $C$ values the capping of the initial P($A$, $B$, $C_c$) along the <111> directions is continued to yield a cubo-octahedral polyhedron of truncated octahedral type as discussed in Sec. 2.2.2 which details all facet shapes, edges, corner coordinates, etc.

**Core $C$ range** of P($A$, -, $C$) where with (86)

$$C \leq C_d \qquad\qquad C/A \leq 1 \qquad\qquad (90)$$

For these $C$ values, the P($A$, -, $C$) polyhedron becomes octahedral and exhibits only {111} facets, see Fig. 3. It is structurally identical with P($A = C$, -, $C$) = P(-, -, $C$) as discussed in Sec. 2.2.4 above and in Sec. 2.1 which details all facet shapes, edges, corner coordinates, etc.

## 2.2.4.3. P($A$, $B$, $C$) by truncating rhombic P(-, $B$, -)

The initial polyhedron **P($A$, $B$, -) with $B \leq A$** is isomorphous to rhombohedral P($A = B$, $B$, -) = P(-, $B$, -) discussed Sec. 2.1. As a result, imposing constraints of a generic octahedral polyhedron, P(-, -, $C$), can yield a rhombo-octahedral polyhedron P(-, $B$, $C$). This requires, according to the discussion above with (58), $C$ values below $C_a = 3/2\, B$. Here we distinguish three different ranges of parameter $C$, defined by separating values $C_a \geq C_c$, where with (58) and (65).

$$C_a(A, B) = 3/2\, B \qquad\qquad (91)$$
$$C_c(A, B) = B \qquad\qquad (92)$$

The ranges are illustrated in Fig. 14 for the rhombohedral polyhedron P(-, $B$, $C_a$).

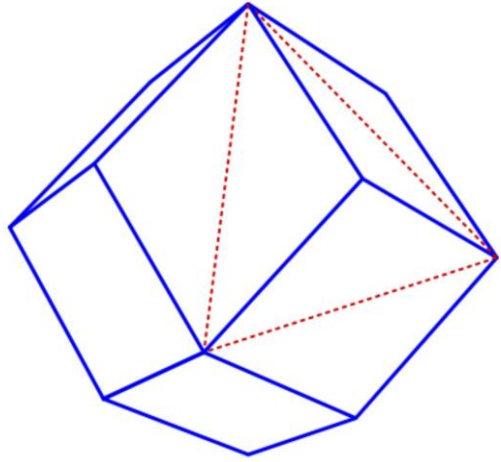

**Figure 14.** Sketch of rhombohedral polyhedron P(-, $B$, $C_a$) ($B/A = 1$, $C/B = 1.5$) with front facets in blue and back facets in black. The cut along a {111} facet corresponding to octahedral P(-, $B$, $C_c$) ($B/A = 1$, $C/B = 1$, red dashed triangle) is added for illustration.

**Outer $C$ range** of P(-, $B$, $C$) where with (91)

$$C \geq C_a \qquad\qquad C/B \geq 3/2 \qquad\qquad (93)$$

For these $C$ values the polyhedron becomes rhombohedral and does not exhibit any {111} facets, see Fig. 14. It is structurally identical with P(-, $B$, $C_a$) = P(-, $B$, -) as discussed above and in Sec. 2.2.3 which details all facet shapes, edges, corner coordinates, etc.

**Central $C$ range** of P(-, $B$, $C$) where with (91), (92)

$$C_c \leq C \leq C_a \qquad\qquad 1 \leq C/B \leq 3/2 \qquad\qquad (94)$$

For these $C$ values the initial P(-, $B$, $C_a$) polyhedron is capped at its <111> corners forming {111} in addition to {100} facets and yielding a rhombo-octahedral polyhedron as discussed in Sec. 2.2.3 which details all facet shapes, edges, corner coordinates, etc.

**Core $C$ range** of P(-, $B$, $C$) where with (92)

$$C \leq C_c \qquad\qquad C/B \leq 1 \qquad\qquad (95)$$

For these $C$ values, the P(-, $B$, $C$) polyhedron becomes octahedral and exhibits only {111} facets, see Fig. 3. It is structurally identical with P(-, $B = C$, $C$) = P(-, -, $C$) as discussed in Sec. 2.2.4 above and in Sec. 2.1 which details all facet shapes, edges, corner coordinates, etc.

## 2.3. Polyhedra of General Facet Families {*abc*}

Going beyond the discussion in Sec. 2.2.4, we consider general compact polyhedra of central **$O_h$ symmetry** confined by facets appearing as parallel pairs with distance $Q$ between them. The orientation of the facets is described by facet normal vectors $\underline{e}$, originating from all $O_h$ symmetry operations applied to generating vectors $\underline{e}_o$ where

$$\underline{e}_o = (a, b, c)/w, \qquad w = \sqrt{(a^2 + b^2 + c^2)} \tag{96}$$

with $\underline{e}_o$ described in Cartesian coordinates $a, b, c$ in arbitrary units relative to the polyhedral center. Due to the overall $O_h$ symmetry, assuming $a \geq b \geq c \geq 0$ covers all cases for each generating vector $\underline{e}_o$. For $a > b > c > 0$ this yields 48 facets with facet normal vectors defining a family <*abc*> of directions $\underline{e}_{<abc>}$ where

$$\begin{aligned}\underline{e}_{<abc>} = \quad & (\pm a, \pm b, \pm c)/w, \quad (\pm a, \pm c, \pm b)/w, \quad (\pm b, \pm a, \pm c)/w, \\ & (\pm b, \pm c, \pm a)/w, \quad (\pm c, \pm a, \pm b)/w, \quad (\pm c, \pm b, \pm a)/w \end{aligned} \tag{97}$$

If any of the values of $a, b, c$ coincide, the number of facet normal vectors will decrease to yield 24, 12, 8, or 6 facets as discussed below. (Note that in the following we use notations <*abc*> (or <*hkl*>) to describe whole families of directions while (*abc*) or (*hkl*) refer to specific directions. Families of facet orientations for given $\underline{e}_{<abc>}$ will be denoted {*abc*} according to a nomenclature introduced for Miller indices [10].) Corresponding compact polyhedra with several generating vectors $\underline{e}_o, \underline{e}'_o, \underline{e}''_o, \ldots$ will be denoted by **P($Q$, {*abc*}; $Q'$, {*a'b'c'*}; $Q''$, {*a"b"c"*}; …)**. While this general case may be quite complex we restrict ourselves in the following to polyhedra **P($Q$, {*abc*})** with only one vector $\underline{e}_o$ generating its family $\underline{e}_{<abc>}$ which can be treated analytically.

As a result of the overall $O_h$ symmetry of the P($Q$, {*abc*}) polyhedron, its corners can appear only along selected directions $\underline{c}_{<hkl>}$ from the center, i.e. only for

$$\underline{c}_{<100>} \quad = (\pm 1, 0, 0), \qquad = (0, \pm 1, 0), \qquad = (0, 0, \pm 1) \tag{98a}$$

$$\underline{c}_{<110>} \quad = 1/\sqrt{2}\,(\pm 1, \pm 1, 0), \quad = 1/\sqrt{2}\,(\pm 1, 0, \pm 1), \quad = 1/\sqrt{2}\,(0, \pm 1, \pm 1) \tag{98b}$$

$$\underline{c}_{<111>} \quad = 1/\sqrt{3}\,(\pm 1, \pm 1, \pm 1) \tag{98c}$$

This yields possible corner vectors

$$\underline{V}_{<hkl>} = p\,\underline{c}_{<hkl>} \tag{99}$$

where $\underline{V}_{<hkl>}$ must point to several joining <*abc*> facet planes which requires

$$\underline{V}_{<hkl>}\,\underline{e}_{<abc>} = p\,(\underline{c}_{<hkl>}\,\underline{e}_{<abc>}) = Q/2 \qquad p = (Q/2)\,/\,(\underline{c}_{<hkl>}\,\underline{e}_{<abc>}) \tag{100}$$

This results in corner lengths $V_{<hkl>}$ as the smallest values of $p$ for all $\underline{e}_{<abc>}$ with $(\underline{c}_{<hkl>}\,\underline{e}_{<abc>}) > 0$, which can be written as

$$V_{<hkl>} = \min_{<abc>} p = (Q/2)\,/\,\max_{<abc>}(\underline{c}_{<hkl>}\,\underline{e}_{<abc>}) \tag{101}$$

For corner directions $\underline{c}_{<100>}$ we obtain with (98a) and $a \geq b \geq c \geq 0$

$$\max{}_{<abc>}(\underline{c}_{<100>} \, \underline{e}_{<abc>}) = \max{}_{<abc>}(a, b, c)/w = a/w \tag{102}$$

which yields with (101) 6 possible corners

$$\underline{V}_{<100>} = (Q/2) \, w/a \, \underline{c}_{<100>} \tag{103}$$

For corner directions $\underline{c}_{<110>}$ we obtain with (98b) and $a \geq b \geq c \geq 0$

$$\max{}_{<abc>}(\underline{c}_{<110>} \, \underline{e}_{<abc>}) = 1/\sqrt{2} \max{}_{<abc>}(a + b, a + c, b + c)/w = 1/\sqrt{2} \, (a + b)/w \tag{104}$$

which yields with (101) 12 possible corners

$$\underline{V}_{<110>} = (Q/2) \, w \, \sqrt{2}/(a + b) \, \underline{c}_{<110>} \tag{105}$$

For corner directions $\underline{c}_{<111>}$ we obtain with (98c) and $a \geq b \geq c \geq 0$

$$\max{}_{<abc>}(\underline{c}_{<111>} \, \underline{e}_{<abc>}) = 1/\sqrt{3} \max{}_{<abc>}(\pm a \pm b \pm c)/w = 1/\sqrt{3} \, (a + b + c)/w \tag{106}$$

and hence yielding with (98c) 8 possible corners

$$\underline{V}_{<111>} = (Q/2) \, w \, \sqrt{3}/(a + b + c) \, \underline{c}_{<111>} \tag{107}$$

## 2.3.1. Polyhedra P(Q, {*abc*})

As discussed in Sec. 2.3 above, most general polyhedra **P(Q, {*abc*})** with $a > b > c > 0$ yield 48 facets with facet normal vectors defining a family {*abc*} where

$$\begin{aligned}\underline{e}_{<abc>} = \quad &(\pm a, \pm b, \pm c)/w, \quad (\pm a, \pm c, \pm b)/w, \quad (\pm b, \pm a, \pm c)/w, \\ &(\pm b, \pm c, \pm a)/w, \quad (\pm c, \pm a, \pm b)/w, \quad (\pm c, \pm b, \pm a)/w \\ w = &\sqrt{(a^2 + b^2 + c^2)}\end{aligned} \tag{97}$$

With (103), (105), (107) this leads to 26 corners described by vectors

$$\underline{V}_{<100>} = (Q/2) \, w/a \, \underline{c}_{<100>} \tag{103}$$

$$\underline{V}_{<110>} = (Q/2) \, w \, \sqrt{2}/(a + b) \, \underline{c}_{<110>} \tag{105}$$

$$\underline{V}_{<111>} = (Q/2) \, w \, \sqrt{3}/(a + b + c) \, \underline{c}_{<111>} \tag{107}$$

According to Eulers polyhedron rule

$$N_c + N_f - N_e = 2 \tag{108}$$

where $N_c$, $N_f$, $N_e$ are the numbers of corners, facets, and edges, respectively, of a convex polyhedron. This yields polyhedra with 72 edges.

A detailed analysis of the polyhedral geometry shows that all 48 facets are of triangular shape with two sets of 24 identical triangles each. The facets of the second set are obtained as mirror images of corresponding facets of the first set. Each facet is described by three edges, connecting <100> with adacent <110>, <100> with <111>, and <110> with <111> corners. The corresponding edge lengths $d_{<100><110>}$, $d_{<100><111>}$, and $d_{<110><111>}$ are found to yield

$$d_{<100><110>} = (Q/2)\, w\, \sqrt{(a^2 + b^2)} / [a\,(a + b)] \tag{109a}$$

$$d_{<100><111>} = (Q/2)\, w\, \sqrt{[2a^2 + (b + c)^2]} / [a\,(a + b + c)] \tag{109b}$$

$$d_{<110><111>} = (Q/2)\, w\, \sqrt{[(a + b)^2 + 2c^2]} / [(a + b)\,(a + b + c)] \tag{109c}$$

Further, the area of each facet is given by $F_{\{abc\}}$ with

$$F_{\{abc\}} = (Qw)^2\, w / [8a\,(a + b)\,(a + b + c)] \tag{110}$$

As a result, the total facet surface $F_{\text{surf}}(\{abc\})$ of the polyhedron is given by

$$F_{\text{surf}}(\{abc\}) = 48 F_{\{abc\}} = 6\,(Qw)^2\, w / [a\,(a + b)\,(a + b + c)] \tag{111}$$

and the volume $V_{\text{tot}}(\{abc\})$ of the polyhedron is given by

$$V_{\text{tot}}(\{abc\}) = (Q/6)\, F_{\text{surf}}(\{abc\}) = (Qw)^3 / [a\,(a + b)\,(a + b + c)] \tag{112}$$

Fig. 15 illustrates the most general polyhedron $P(Q, \{abc\})$ for $a = 7$, $b = 3$, $c = 1$.

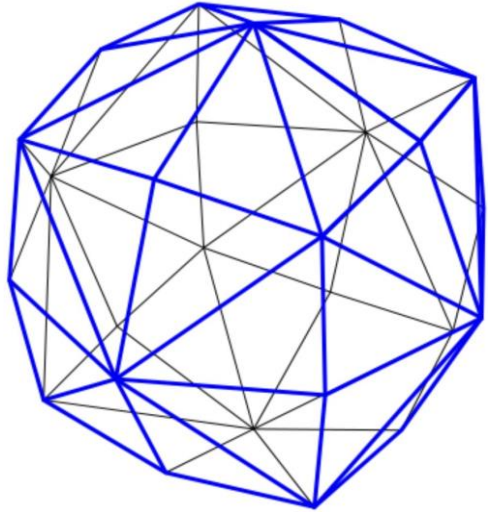

**Figure 15.** Sketch of polyhedron of type $\{abc\}$ with $a = 7$, $b = 3$, c = 1 with front facets in blue and back facets in black.

## 2.3.2. Polyhedra P(Q, {*ab*0})

Polyhedra **P(*Q*, {*ab*0})** are represented by $a > b > c = 0$ and corresponding facet normal vectors, defining a family {*ab*0}, are obtained from the general set (97) reducing to 24 vectors where

$$\underline{e}_{<ab0>} = \quad (\pm a, \pm b, 0)/w, \quad (\pm b, \pm a, 0)/w, \quad (\pm a, 0, \pm b)/w,$$
$$\quad\quad\quad (\pm b, 0, \pm a)/w, \quad (0, \pm a, \pm b)/w, \quad (0, \pm b, \pm a)/w \quad (113)$$
$$w = \sqrt{(a^2 + b^2)}$$

yielding 24 facets where the possible corner directions $\underline{c}_{<hkl>}$ given by (2.3) remain the same. With (103), (107) this leads to 14 corners described by vectors

$$\underline{V}_{<100>} = (Q/2)\, w/a\; \underline{c}_{<100>} \quad (114)$$

$$\underline{V}_{<111>} = (Q/2)\, w\, \sqrt{3}/(a + b)\; \underline{c}_{<111>} \quad (115)$$

For this polyhedron type all possible <110> corners of the general case (105) with

$$\underline{V}_{<110>} = (Q/2)\, w\, \sqrt{2}/(a + b)\; \underline{c}_{<110>} \quad (116)$$

are located in the center between corresponding adjacent <111> corners. For example, vector $\underline{V}_{(110)}$ lies between $\underline{V}_{(111)}$ and $\underline{V}_{(11\text{-}1)}$ according to

$$\underline{V}_{(110)} = (Q/2)\, w\,/(a + b)\, (1, 1, 0) = 1/2\, [\underline{V}_{(111)} + \underline{V}_{(11\text{-}1)}] \quad (117)$$

As a result, these <110> "corners" become midpoints of edges between <111> corners reducing the total number of polyhedral corners to 14. Further, pairs of facet triangles joining along lines connecting <100> with <110> "corners" belong to identical facet normal vectors $\underline{e}_{<ab0>}$ to combine as larger facets of isosceles shape with two edges connecting <100> with <110> and one connecting two adjacent <111> corners, see Fig. 16a, yielding edge lengths

$$d_{<100><111>} = (Q/2)\, w\, \sqrt{[2a^2 + b^2]}\, /\, [a\,(a + b)] \quad (118\text{a})$$

$$d_{<111><111>} = Q\, w\, /\, (a + b) \quad (118\text{b})$$

$$d_{<100><111>} < d_{<111><111>} \quad (118\text{c})$$

The area of each facet is given by $F_{\{ab0\}}$ with

$$F_{\{ab0\}} = (Qw)^2\, w\, /\, [4a\,(a + b)^2] \quad (119)$$

Thus, the total facet surface $F_{\text{surf}}(\{ab0\})$ of the polyhedron is given by

$$F_{\text{surf}}(\{ab0\}) = 24 F_{\{ab0\}} = 6(Qw)^2\, w\, /\, [a\,(a + b)^2] \quad (120)$$

and the volume $V_{\text{tot}}(\{ab0\})$ of the polyhedron is given by

$$V_{\text{tot}}(\{ab0\}) = (Q/6)\, F_{\text{surf}}(\{ab0\}) = (Qw)^3\, /\, [a\,(a + b)^2] \quad (121)$$

With polyhedron P(*Q*, {*ab*0}) yielding 24 facets and 14 corners the number of its polyhedral edges amounts to 36 according to (108). In fact, the general shape of a polyhedron P(*Q*, {*ab*0}) can be characterized qualitatively as a cube whose six surface sides are complemented by identical square pyramids. Fig.16b illustrates a polygon P(Q, {ab0}) for parameters $a = 7$, $b = 4$, $c = 0$.

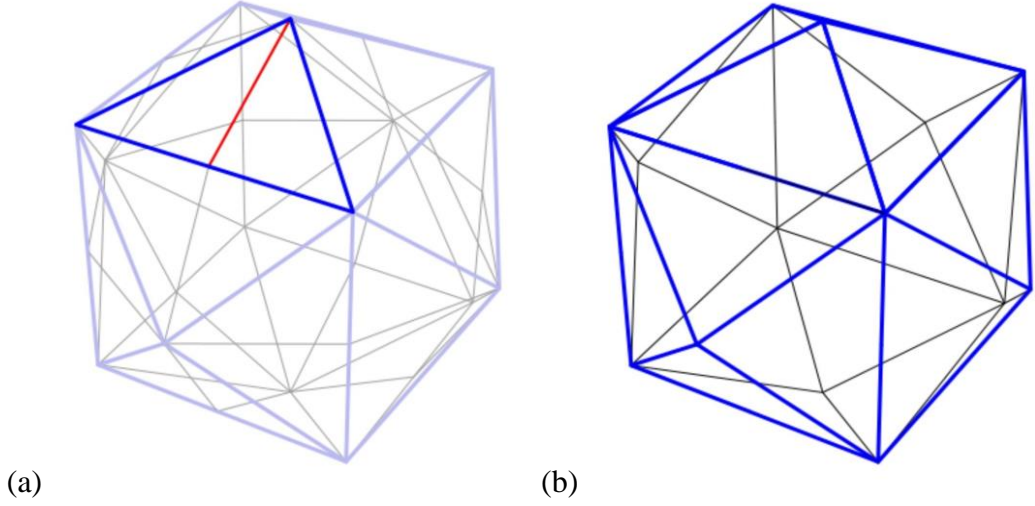

(a) (b)

**Figure 16.** Sketch of polyhedron of type {ab0} with $a = 7$, $b = 4$, c = 0. (a) Polyhedron sketch with front facets in light blue and back facets in gray, two joining coplanar triangular facets are emphasized in dark blue with the separating <100>-<110> line in red; (b) full polyhedron sketch with front facets in blue and back facets in black, see text.

### 2.3.3. Polyhedra P(Q, {abb})

Polyhedra **P(Q, {abb})** are represented by $a > b = c > 0$ and corresponding facet normal vectors, defining a family {abb}, are obtained from the general set (97) reducing to 24 vectors where

$$\underline{e}_{<abb>} = (\pm a, \pm b, \pm b)/w, \quad (\pm b, \pm a, \pm b)/w, \quad (\pm b, \pm b, \pm a)/w \tag{122}$$
$$w = \sqrt{(a^2 + 2b^2)}$$

yielding 24 facets where the possible corner directions $\underline{c}_{<hkl>}$ given by (2.3) remain the same. With (103), (105), (107) this leads to 26 corners described by vectors

$$\underline{V}_{<100>} = (Q/2)\, w/a\, \underline{c}_{<100>} \tag{123}$$

$$\underline{V}_{<110>} = (Q/2)\, w\, \sqrt{2}/(a + b)\, \underline{c}_{<110>} \tag{124}$$

$$\underline{V}_{<111>} = (Q/2)\, w\, \sqrt{3}/(a + 2b)\, \underline{c}_{<111>} \tag{125}$$

For this polyhedron type, pairs of facet triangles joining along lines connecting <100> with <111> corners belong to identical facet normal vectors $\underline{e}_{<abb>}$ to combine as larger facets of quadrangle shape (kite shaped) with two edges connecting <100> with <110> and two connecting <111> with <110> corners, see Fig. 17a, yielding edge lengths

$$d_{<100><110>} = (Q/2)\, w\, \sqrt{(a^2 + b^2)} / [a\,(a + b)] \quad (126a)$$

$$d_{<111><110>} = (Q/2)\, w\, \sqrt{[(a + b)^2 + 2b^2]} / [(a + b)\,(a + 2b)] \quad (126b)$$

$$d_{<111><110>} < d_{<100><110>} \quad (126c)$$

The area of each facet is given by $F_{\{abb\}}$ with

$$F_{\{abb\}} = (Qw)^2\, w / [4a\,(a + b)\,(a + 2b)] \quad (127)$$

Thus, the total facet surface $F_{\text{surf}}(\{abb\})$ of the polyhedron is given by

$$F_{\text{surf}}(\{abb\}) = 24 F_{\{abb\}} = 6(Qw)^2\, w / [a\,(a + b)\,(a + 2b)] \quad (128)$$

and the volume $V_{\text{tot}}(\{abb\})$ of the polyhedron is given by

$$V_{\text{tot}}(\{abb\}) = (Q/6)\, F_{\text{surf}}(\{abb\}) = (Qw)^3 / [a\,(a + b)\,(a + 2b)] \quad (129)$$

With polyhedron P($Q$, $\{abb\}$) yielding 24 facets and 26 corners the number of its polyhedral edges amounts to 48 according to (108). Fig.17b illustrates a polyhedron P($Q$, $\{abb\}$) for parameters $a = 9$, $b = c = 4$.

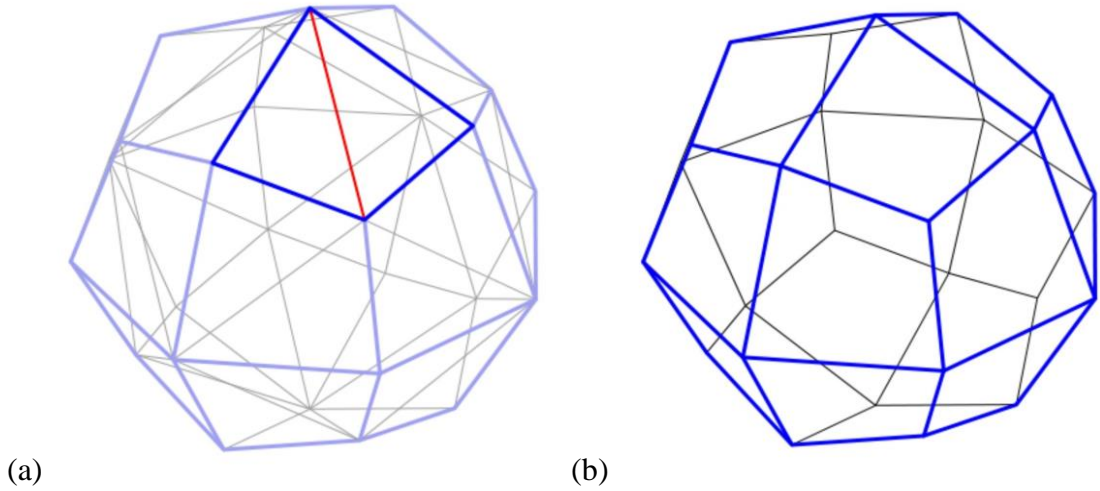

(a)          (b)

**Figure 17.** Sketch of polyhedron of type $\{abb\}$ with $a = 9$, $b = c = 4$. (a) Polyhedron sketch with front facets in light blue and back facets in gray, two joining coplanar triangular facets are emphasized in dark blue with the separating <100>-<111> line in red; (b) full polyhedron sketch with front facets in blue and back facets in black, see text.

## 2.3.4. Polyhedra P(Q, {*aac*})

Polyhedra **P(*Q*, {*aac*})** are represented by $a = b > c > 0$ and corresponding facet normal vectors, defining a family {*aac*}, are obtained from the general set (97) reducing to 24 vectors where

$$\underline{e}_{<aac>} = (\pm a, \pm a, \pm c)/w, \quad (\pm a, \pm c, \pm a)/w, \quad (\pm c, \pm a, \pm a)/w \quad (130)$$
$$w = \sqrt{(2a^2 + c^2)}$$

yielding 24 facets where the possible corner directions $\underline{c}_{<hkl>}$ given by (2.3) remain the same. With (103), (107) this leads to 14 corners described by vectors

$$\underline{V}_{<100>} = (Q/2)\, w/a\, \underline{c}_{<100>} \quad (131)$$
$$\underline{V}_{<111>} = (Q/2)\, w\, \sqrt{3}/(2a + c)\, \underline{c}_{<111>} \quad (132)$$

For this polyhedron type all possible <110> corners of the general case (105) with

$$\underline{V}_{<110>} = (Q/2)\, w/(\sqrt{2}a)\, \underline{c}_{<110>} \quad (133)$$

are located in the center between corresponding adjacent <100> corners. For example, vector $\underline{V}_{(110)}$ lies between $\underline{V}_{(100)}$ and $\underline{V}_{(010)}$ according to

$$\underline{V}_{(110)} = (Q/2)\, w/(2a)\, (1, 1, 0) = 1/2\, [\underline{V}_{(100)} + \underline{V}_{(010)}] \quad (134)$$

As a result, these <110> "corners" become midpoints of edges between <100> corners reducing the total number of polyhedral corners to 14. Further, pairs of facet triangles joining along lines connecting <110> with <111> "corners" belong to identical facet normal vectors $\underline{e}_{<aac>}$ to combine as larger facets of isosceles shape with two edges connecting <100> with <111> and one connecting two adjacent <100> corners, see Fig. 18a, yielding edge lengths

$$d_{<100><111>} = (Q/2)\, w\, \sqrt{[2a^2 + (a + c)^2]} / [a\,(2a + c)] \quad (135a)$$
$$d_{<100><100>} = Q\, w / \sqrt{2}a \quad (135b)$$
$$d_{<100><111>} < d_{<100><100>} \quad (135c)$$

The area of each facet is given by $F_{\{aac\}}$ with

$$F_{\{aac\}} = (Qw)^2\, w / [8a^2\,(2a + c)] \quad (136)$$

As a result, the total facet surface $F_{\text{surf}}(\{aac\})$ of the polyhedron is given by

$$F_{\text{surf}}(\{aac\}) = 24 F_{\{aac\}} = 3\,(Qw)^2\, w / [a^2\,(2a + c)] \quad (137)$$

and the volume $V_{\text{tot}}(\{aac\})$ of the polyhedron is given by

$$V_{\text{tot}}(\{aac\}) = (Q/6)\, F_{\text{surf}}(\{aac\}) = (Qw)^3 / [2a^2\,(2a + c)] \quad (138)$$

With polyhedron P(*Q*, {*aac*}) yielding 24 facets and 14 corners the number of its polyhedral edges amounts to 36 according to (108). Fig.18b illustrates a polyhedron P(*Q*, {*aac*}) for parameters $a = b = 8$, $c = 3$.

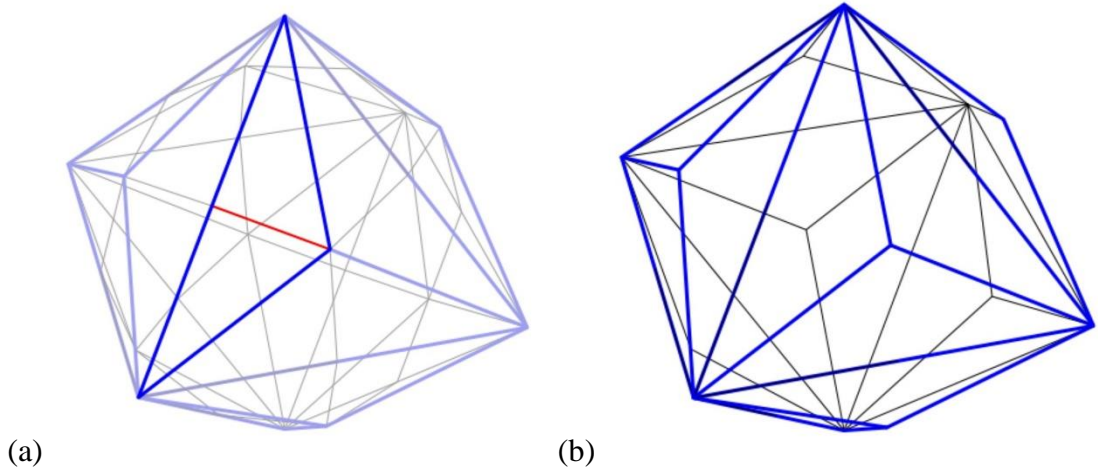

(a) (b)

**Figure 18.** Sketch of polyhedron of type {*aac*} with $a = b = 8$, $c = 3$. (a) Polyhedron sketch with front facets in light blue and back facets in gray, two joining coplanar triangular facets are emphasized in dark blue with the separating <110>-<111> line in red; (b) full polyhedron sketch with front facets in blue and back facets in black, see text.

### 2.3.5. Polyhedra P(Q, {*aaa*})

Polyhedra **P(*Q*, {*aaa*})** represented by $a = b = c > 0$ can be considered as special cases of polyhedra P(*Q*, {*aac*}) discussed in Sec. 2.3.4. Corresponding facet normal vectors, defining a family {*aaa*}, are obtained from the general set (97) reducing to 8 vectors where

$$\underline{e}_{<aaa>} = (\pm a, \pm a, \pm a)/w, \qquad w = \sqrt{3}\, a \tag{139}$$

yielding 8 facets where the possible corner directions $\underline{c}_{<hkl>}$ given by (2.3) remain the same. With (103) this leads to 6 corners described by vectors

$$\underline{V}_{<100>} = (Q/2)\, w/a\, \underline{c}_{<100>} \tag{140}$$

For this polyhedron type all possible <110> corners of the general case (105) with

$$\underline{V}_{<110>} = (Q/2)\, w/(\sqrt{2}a)\, \underline{c}_{<110>} \tag{141}$$

are located in the center between two corresponding adjacent <100> corners. For example, vector $\underline{V}_{(110)}$ lies between $\underline{V}_{(100)}$ and $\underline{V}_{(010)}$ according to

$$\underline{V}_{(110)} = (Q/2)\, w/(2a)\, (1, 1, 0) = 1/2\, [\underline{V}_{(100)} + \underline{V}_{(010)}] \tag{142}$$

Further, all possible <111> corners of the general case (107) with

$$\underline{V}_{<111>} = (Q/2)\, w / (\sqrt{3}\, a)\, \underline{c}_{<111>} \tag{143}$$

lie in the center between three corresponding adjacent <100> corners. For example, vector $\underline{V}_{(111)}$ lies between $\underline{V}_{(100)}$, $\underline{V}_{(010)}$, and $\underline{V}_{(001)}$ according to

$$\underline{V}_{<111>} = (Q/2)\, w / (\sqrt{3}\, a)\, \underline{c}_{<111>} = 1/3\, [\underline{V}_{(100)} + \underline{V}_{(010)} + \underline{V}_{(001)}] \tag{144}$$

As a result, <110> "corners" become midpoints of edges between pairs of <100> corners and <111> "corners" lie in the center of planes through three <100> corners, see Fig. 19a. Altogether, this leads to facets of equilateral triangular shape yielding edge lengths

$$d_{<100><100>} = Q\,w\,/\,(\sqrt{2}a) \qquad (145)$$

The area of each facet is given by $F_{\{aaa\}}$ with

$$F_{\{aaa\}} = (Qw)^2\,w\,/\,(8a^3) \qquad (146)$$

As a result, the total facet surface $F_{\text{surf}}(\{aaa\})$ of the polyhedron is given by

$$F_{\text{surf}}(\{aaa\}) = 8F_{\{aaa\}} = (Qw)^2\,w\,/\,a^3 \qquad (147)$$

and the volume $V_{\text{tot}}(\{aaa\})$ of the polyhedron is given by

$$V_{\text{tot}}(\{aaa\}) = (Q/6)\,F_{\text{surf}}(\{aaa\}) = (Qw)^3\,/\,6a^3 \qquad (148)$$

With polyhedron P($Q$, $\{aaa\}$) yielding 8 facets and 6 corners the number of its polyhedral edges amounts to 12 according to (108). In fact, the general shape of a polyhedron P($Q$, $\{aaa\}$) can be characterized qualitatively as a regular octahedron discussed in Sec. 2.1. Hence we may write formally P($Q$, $\{aaa\}$) = P(-, -, $C = Q\sqrt{3}$). Fig.19b illustrates a polyhedron P($Q$, $\{aaa\}$) for parameters $a = b = c = 1$.

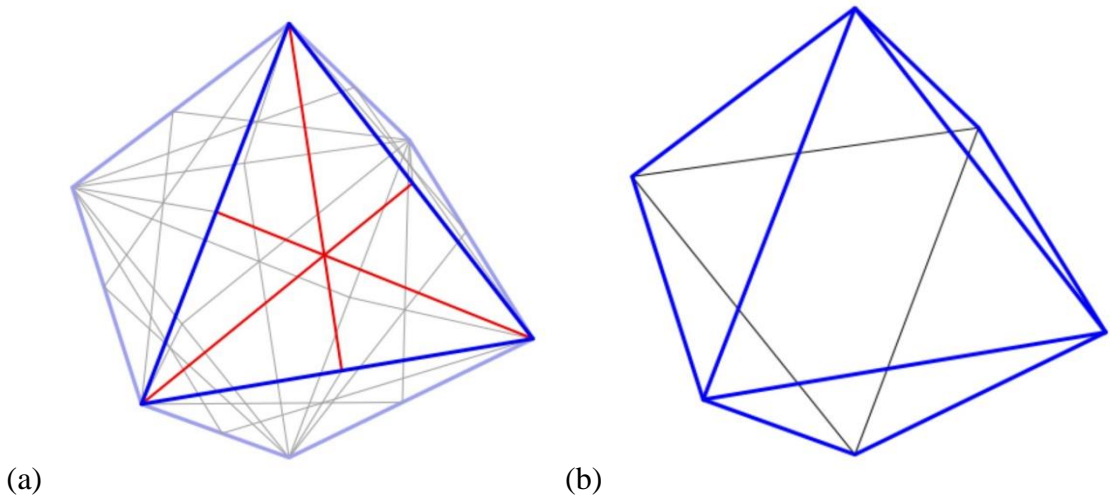

(a) (b)

**Figure 19.** Sketch of polyhedron of type $\{aaa\}$ with $a = b = c = 1$. (a) Polyhedron sketch with front facets in light blue and back facets in gray, six joining coplanar triangular facets are emphasized in dark blue with the separating <100>-<111> lines in red; (b) full polyhedron sketch with front facets in blue and back facets in black, see text.

## 2.3.6. Polyhedra P(Q, {*aa*0})

Polyhedra **P(*Q*, {*aa*0})** represented by $a = b > c = 0$ can be considered as special cases of polyhedra P(*Q*, {*ab*0}) discussed in Sec. 2.3.2. Corresponding facet normal vectors, defining a family {*aa*0}, are obtained from the general set (97) reducing to 12 vectors where

$$\underline{e}_{<aa0>} = (\pm a, \pm a, 0)/w, \quad (\pm a, 0, \pm a)/w, \quad (0, \pm a, \pm a)/w \quad (149)$$
$$w = \sqrt{2}\, a$$

yielding 12 facets where the possible corner directions $\underline{c}_{<hkl>}$ given by (2.3) remain the same. With (103), (107) this leads to 14 corners described by vectors

$$\underline{V}_{<100>} = (Q/2)\, w/a\, \underline{c}_{<100>} \quad (150)$$
$$\underline{V}_{<111>} = (Q/2)\, w\, \sqrt{3}/(2a)\, \underline{c}_{<111>} \quad (151)$$

For this polyhedron type all possible <110> corners of the general case (105) with

$$\underline{V}_{<110>} = (Q/2)\, w\, \sqrt{2}/(2a)\, \underline{c}_{<110>} \quad (152)$$

are located in the center between corresponding adjacent <111> as well as <100> corners. For example, vector $\underline{V}_{(110)}$ lies between $\underline{V}_{(111)}$ and $\underline{V}_{(11\text{-}1)}$ according to

$$\underline{V}_{(110)} = (Q/4)\, w\, /a\, (1, 1, 0) = 1/2\, [\underline{V}_{(111)} + \underline{V}_{(11\text{-}1)}] \quad (153)$$

but also between $\underline{V}_{(100)}$ and $\underline{V}_{(010)}$ according to

$$\underline{V}_{(110)} = (Q/4)\, w\, /a\, (1, 1, 0) = 1/2\, [\underline{V}_{(100)} + \underline{V}_{(010)}] \quad (154)$$

As a result, the four facet triangles sharing each <110> "corner" belong to identical facet normal vectors $\underline{e}_{<aa0>}$ to combine as larger facets of rhombic shape with four edges connecting <100> with <111> corners, see Fig. 20a, yielding edge lengths

$$d_{<100><111>} = (Q/4)\, w\, \sqrt{3}\, /\, a \quad (155)$$

The area of each facet is given by $F_{\{aa0\}}$ with

$$F_{\{aa0\}} = (Qw)^2\, w\, /\, 8a^3 \quad (156)$$

Thus, the total facet surface $F_{\text{surf}}(\{aa0\})$ of the polyhedron is given by

$$F_{\text{surf}}(\{aa0\}) = 12 F_{\{aa0\}} = 3(Qw)^2\, w\, /\, 2a^3 \quad (157)$$

and the volume $V_{\text{tot}}(\{aa0\})$ of the polyhedron is given by

$$V_{\text{tot}}(\{aa0\}) = (Q/6)\, F_{\text{surf}}(\{aa0\}) = (Qw)^3\, /\, 4a^3 \quad (158)$$

With polyhedron P(*Q*, {*aa*0}) yielding 12 facets and 14 corners the number of its polyhedral edges amounts to 24 according to (108). In fact, the general shape of a polyhedron P(*Q*, {*aa*0}) can be characterized qualitatively as a regular rhombic dodecahedron discussed in Sec. 2.1. Hence we may write formally P(*Q*, {*aa*0}) = P(-, *B* = *Q* $\sqrt{2}$, -). Fig.20b illustrates a polyhedron P(*Q*, {*aa*0}) for parameters $a = b = 1$, $c = 0$.

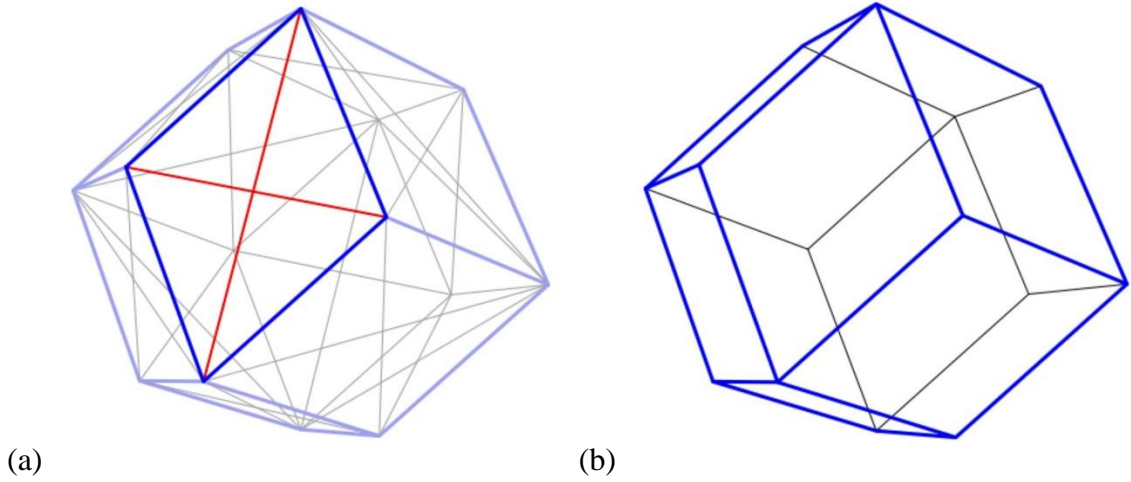

(a) (b)

**Figure 20.** Sketch of polyhedron of type {$aa0$} with $a = b = 1$, $c = 0$. (a) Polyhedron sketch with front facets in light blue and back facets in gray, four joining coplanar triangular facets are emphasized in dark blue with the separating <110>-<100> and <110>-<111> lines in red; (b) full polyhedron sketch with front facets in blue and back facets in black, see text.

### 2.3.7. Polyhedra P(Q, {$a00$})

Polyhedra **P**($Q$, {$a00$}) represented by $a > b = c = 0$ can be considered as special cases of polyhedra **P**($Q$, {$ab0$}) discussed in Sec. 2.3.2. Corresponding facet normal vectors, defining a family {$a00$}, are obtained from the general set (97) reducing to 6 vectors where

$$\underline{e}_{<a00>} = (\pm a, 0, 0)/w, \quad (0, \pm a, 0)/w, \quad (0, 0, \pm a)/w \quad (159)$$
$$w = a$$

yielding 6 facets where the possible corner directions $\underline{c}_{<hkl>}$ given by (2.3) remain the same. With (107) this leads to 8 corners described by vectors

$$\underline{V}_{<111>} = (Q/2)\sqrt{3}\,\underline{c}_{<111>} \quad (160)$$

For this polyhedron type all possible <110> corners of the general case (105) with

$$\underline{V}_{<110>} = (Q/2)\sqrt{2}\,\underline{c}_{<110>} \quad (161)$$

are located in the center between corresponding adjacent <111> corners. For example, vector $\underline{V}_{(110)}$ lies between $\underline{V}_{(111)}$ and $\underline{V}_{(11\text{-}1)}$ according to

$$\underline{V}_{(110)} = (Q/2)(1, 1, 0) = 1/2\,[\underline{V}_{(111)} + \underline{V}_{(11\text{-}1)}] \quad (162)$$

As a result, these <110> "corners" become midpoints of edges between <111> corners. Further, all possible <100> corners of the general case (103) with

$$\underline{V}_{<100>} = (Q/2)\,\underline{c}_{<100>} \quad (163)$$

are located in the center of four <111> corners. For example, vector $\underline{V}_{(100)}$ lies in the center of $\underline{V}_{(111)}$, $\underline{V}_{(11\text{-}1)}$, $\underline{V}_{(1\text{-}11)}$, and $\underline{V}_{(1\text{-}1\text{-}1)}$ according to

$$\underline{V}_{<100>} = (Q/2)\, \underline{c}_{<100>} = 1/4\, [\underline{V}_{(111)} + \underline{V}_{(11\text{-}1)} + \underline{V}_{(1\text{-}11)} + \underline{V}_{(1\text{-}1\text{-}1)}] \tag{164}$$

As a result, the 8 facet triangles sharing each <100> "corner" belong to identical facet normal vectors $\underline{e}_{<a00>}$ to combine as larger facets of square shape with four edges connecting <111> with <111> corners, see Fig. 21a, yielding edge lengths

$$d_{<111><111>} = Q \tag{165}$$

The area of each facet is given by $F_{\{a00\}}$ with

$$F_{\{a00\}} = Q^2 \tag{166}$$

Thus, the total facet surface $F_{\text{surf}}(\{a00\})$ of the polyhedron is given by

$$F_{\text{surf}}(\{a00\}) = 6F_{\{a00\}} = 6Q^2 \tag{167}$$

With polyhedron P($Q$, $\{a00\}$) yielding 6 facets and 8 corners the number of its polyhedral edges amounts to 12 according to (108). In fact, the general shape of a polyhedron P($Q$, $\{a00\}$) can be characterized qualitatively as a generic cube discussed in Sec. 2.1. Hence we may write formally P($Q$, $\{a00\}$) = P($A = Q$, -, -). Fig. 21b illustrates a polyhedron P($Q$, $\{a00\}$) for parameters $a = 1$, $b = c = 0$.

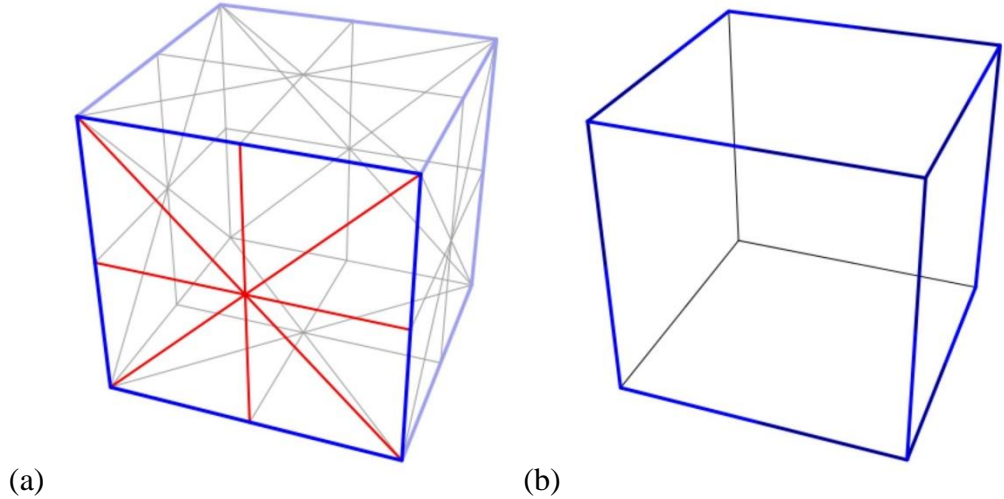

(a)     (b)

**Figure 21.** Sketch of polyhedron of type $\{a00\}$ with $a > b = c = 0$. (a) Polyhedron sketch with front facets in light blue and back facets in gray, eight joining coplanar triangular facets are emphasized in dark blue with the separating <100>-<110> and <100>-<111> lines in red; (b) full polyhedron sketch with front facets in blue and back facets in black, see text.

## 3. Conclusions

The present theoretical analysis gives a full account of the shape and structure of compact polyhedra with cubic $O_h$ symmetry confined by planar facets where. Here we focus on facets reflecting normal directions of families {100}, {110}, and {111} in Cartesian coordinates and located at different distances from the polyhedral center. The structure evaluation identifies different types of generic NPs, which serve for the definition of general polyhedral NPs as intersections of corresponding generic NPs. This allows a characterization of polyhedra according to only three polyhedral parameters *A*, *B*, *C*. These are connected with polyhedron diameters along corresponding facet normal directions and allow an easy classification of all corresponding polyhedra. Further, structural properties of compact polyhedra of $O_h$ symmetry, confined by facets with normal vectors of one general {*abc*} family, are shown to be characterized by four polyhedral structure parameters *Q*, *a*, *b*, *c*. Here *Q* represents the polyhedron diameter and *a*, *b*, *c* specify the normal directions of the facets. Detailed structural properties of all polyhedra, such as shape, size, and facet surfaces, are discussed in analytical and numerical detail with visualization of characteristic examples. Altogether, these results allow an easy classification of all corresponding polyhedra which may also be used as a repository for structures of real compact NPs with internal cubic lattice observed by experiment.

# S. Supplementary Information

## S1. Notations of Polyhedra

This work uses two different notations,

    notation 1:        P($A$, $B$, $C$),

    notation 2:        P($Q$, {$abc$}; $Q'$, {$a'b'c'$}; $Q''$, {$a''b''c''$}; …)

to characterize polyhedra with cubic $O_h$ symmetry where the latter includes the former but is more general. The following list gives translations between the notations in all cases where there is equivalence.

| Notation 1 | Notation 2 |
|---|---|
| P($A$, $B$, $C$) | P($Q = A$, {100}; $Q = B/\sqrt{2}$, {110}; $Q = C/\sqrt{3}$, {111};) |
| P($A$, $B$, -) | P($Q = A$, {100}; $Q = B/\sqrt{2}$, {110}) |
| P($A$, -, $C$) | P($Q = A$, {100}; $Q = C/\sqrt{3}$, {111}) |
| P(-, $B$, $C$) | P($Q = B/\sqrt{2}$, {110}; $Q = C/\sqrt{3}$, {111}) |
| P($A$, -, -) | P($Q = A$, {100}) |
| P(-, $B$, -) | P($Q = B/\sqrt{2}$, {110} |
| P(-, -, $C$) | P($Q = C/\sqrt{3}$, {111}) |

## S2. Classification Tables of Polyhedra P($A$, $B$, $C$)

Here we present Tables giving a full classification of $O_h$ symmetry polyhedra P($A$, $B$, $C$) for all combinations of polyhedral parameters $A$, $B$, $C$. This includes polyhedra where one or two parameters define the structure already uniquely and missing parameter values are replaced by "-". Further, isomorphs of a given polyhedron are defined as those which are structurally identical but differ in their notation. As an example, all polyhedra P($A$, $B$, $C$) with $B > 2A$ and $C > 3A$ are isomorphs of the generic polyhedron P($A$, $2A$, $3A$).

### S.2.1. Generic Polyhedra P(*A*, -, -), (-, *B*, -), (-, -, *C*)

| Generic type | Facets | Corners |
|---|---|---|
| Cubic<br>P(*A*, -, -) | {100} 6<br>{110} 0<br>{111} 0 | <100> 0<br><110> 0<br><111> 8 |
| Rhombohedral<br>P(-, *B*, -) | {100} 0<br>{110} 12<br>{111} 0 | <100> 6<br><110> 0<br><111> 8 |
| Octahedral<br>P(-, -, *C*) | {100} 0<br>{110} 0<br>{111} 8 | <100> 6<br><110> 0<br><111> 0 |

**Table 1.** Types and notations of all generic polyhedra P(*A*, -, -), P(-, *B*, -), P(-, -, *C*).

### S.2.2. Non-generic Polyhedra P(*A*, *B*, -), (*A*, -, *C*), (-, *B*, *C*)

| Constraints | MP types | Isomorphs |
|---|---|---|
| *B* ≥ 2*A* | Generic cubic | (*A*, -, -) =<br>(*A*, *B* = 2*A*, -) |
| *A* < *B* < 2*A* | Cubo-rhombic | (*A*, *B*, -) |
| *B* ≤ *A* | Generic rhombohedral | (-, *B*, -) =<br>(*A* = *B*, *B*, -) |

**Table 2.** Constraints and types including isomorphs of polyhedra P(*A*, *B*, -).

| Constraints | MP types | Isomorphs |
|---|---|---|
| $C \geq 3A$ | Generic cubic | $(A, -, -) =$ <br> $(A, -, C = 3A)$ |
| $2A \leq C \leq 3A$ | Cubo-octahedral, truncated cubic | $(A, -, C)$ |
| $C = 2A$ | Cuboctahedral | $(A, -, C = 2A)$, <br> $(A = C/2, -, C)$ |
| $A \leq C \leq 2A$ | Cubo-octahedral, truncated octahedral | $(A, -, C)$ |
| $C \leq A$ | Generic octahedral | $(-, -, C) =$ <br> $(A = C, -, C)$ |

**Table 3.** Constraints and types including isomorphs of polyhedra $P(A, -, C)$.

| Constraints | MP types | Isomorphs |
|---|---|---|
| $C \geq 3/2\ B$ | Generic rhombohedral | $(-, B, -) =$ <br> $(-, B, C = 3/2\ B)$ |
| $B \leq C \leq 3/2\ B$ | Rhombo-octahedral | $(-, B, C)$ |
| $C \leq B$ | Generic octahedral | $(-, -, C) =$ <br> $(-, B = C, C)$ |

**Table 4.** Constraints and types including isomorphs of polyhedra $P(-, B, C)$.

## S.2.3. General non-generic Polyhedra P(*A*, *B*, *C*)

| Constraints 1 | Constraints 2 | MP types | Isomorphs |
|---|---|---|---|
| $B \geq 2A$ | $C \geq 3A$ | Generic cubic | $(A, -, -) =$ $(A, B_a, C_a)$ |
| | $2A \leq C \leq 3A$ | Cubo-octahedral, truncated cubic | $(A, -, C) =$ $(A, B_a, C)$ |
| | $C = 2A$ | Cuboctahedral | $(A, B_a, C)$ |
| | $A \leq C \leq 2A$ | Cubo-octahedral, truncated octahedral | $(A, -, C) =$ $(A, B_a, C)$ |
| | $C \leq A$ | Generic octahedral | $(-, -, K) =$ $(A_a, B_a, C)$ |
| $A \leq B \leq 2A$ | $C \geq 3/2\ B$ | Cubo-rhombic | $(A, B, -) =$ $(A, B, 3/2\ B)$ |
| | $2B - A \leq C \leq 3/2\ B$ | Cubo-rhombo-oct., upper central | $(A, B, C)$ |
| | $B \leq C \leq 2B - A$ | Cubo-rhombo-oct., lower central | $(A, B, C)$ |
| | $A \leq C \leq B$ | Cubo-octahedral, truncated octahedral | $(A, -, C) =$ $(A, C, C)$ |
| | $C \leq A$ | Generic octahedral | $(-, -, C) =$ $(C, C, C)$ |
| $B \leq A$ | $C \geq 3/2\ B$ | Generic rhombohedral | $(-, B, -) =$ $(A_a, B, C_a)$ |
| | $B \leq C \leq 3/2\ B$ | Rhombo-octahedral | $(-, B, C) =$ $(A_a, B, C)$ |
| | $C \leq B$ | Generic octahedral | $(-, -, C) =$ $(A_a, B_a, C)$ |

**Table 5.** Constraints and types including isomorphs of cub(*A*, *B*, *C*) MPs. Polyhedral parameters $A_a$, $B_a$, $C_a$ are defined by (62), (60), (58).

## S.3. Mathematical relationships

In this section we list simple mathematical relationships which can be used to calculate areas and volumes of basic polyhedra. The corresponding formulas for areas F and volumes V have been applied to evaluate facet areas and volumes of all polyhedra P(*A*, *B*, *C*) in Secs. 2.1-4.

### S.3.1. Areas

**Rectangle** with edges $a \neq b$ : $\quad F_{rect} = a\,b$

**Square** with edges $a$ : $\quad F_{square} = a^2$

**Triangle** with edges $a$ : $\quad F_{triangle} = \sqrt{3}/4\, a^2$

**Hexagon**, capped triangle with edges $a \neq b$ : $\quad F_{hex} = \sqrt{3}/4\,[(a + 2b)^2 - 3b^2]$

regular with edges $a$ : $\quad F_{rhex} = 3\sqrt{3}/2\, a^2$

**Octagon**, capped square with edges $a \neq b$ : $\quad F_{oct} = (a + \sqrt{2}\,b)^2 - b^2$

regular with edges $a$ : $\quad F_{oct} = 2(1 + \sqrt{2})^2\, a^2$

### S.3.2. Volumes

**Cube** with edges $a$ : $\quad V_{cube} = a^3$

**Octahedron** with edges $a$ : $\quad V_{oct} = \sqrt{2}/3\, a^3$

**Rhombic dodecahedron** with edges $a$ : $\quad V_{rhomb} = 16/\sqrt{27}\, a^3$

**Pyramid**, triangular base with edges $a$, height $d$ : $\quad V_{tpyr} = \sqrt{3}/12\, a^2\, d$

square base with edges $a$, height $d$ : $\quad V_{spyr} = 1/3\, a^2\, d$

rectangular base with edges $a$, $b$, height $d$ : $\quad V_{rpyr} = 1/3\, a\, b\, d$

**Ridge**, rectangular base with edges $a$, $b$, height $d$,
linear ridge top of length $c$ $\quad V_{rlpyr} = a\, d\,[c/2 + (b - c)/3]$